\documentclass[reprint,amsmath,amssymb,aps,prd]{revtex4-2}
\usepackage{graphicx}
\usepackage{dcolumn}
\usepackage{bm}
\usepackage{siunitx}
\usepackage{multirow}
\usepackage{booktabs}
\usepackage{float}
\usepackage{hyperref}
\usepackage{cleveref}
\hypersetup{colorlinks = true, linkcolor = black, urlcolor = blue, citecolor = blue}

\begin{document}

\title{The radiative decay of scalar glueball from lattice QCD}

\author{Jintao Zou }
\affiliation{  Department of Physics, Hunan Normal University,  Changsha 410081, China }

\author{Long-Cheng Gui \footnote{Corresponding author}} \email{guilongcheng@hunnu.edu.cn}
\affiliation{  Department of Physics, Hunan Normal University, and Key Laboratory of Low-Dimensional Quantum Structures and Quantum Control of Ministry of Education,   Changsha 410081, China }
\affiliation{Synergetic Innovation Center for Quantum Effects and Applications (SICQEA), Hunan Normal University, Changsha 410081, China }

\author{Ying Chen } \email{cheny@ihep.ac.cn}
\affiliation{  Institute of High Energy Physics, Chinese Academy of Sciences, Beijing 100049, People’s Republic of China }

\author{Jian Liang }
\affiliation{Key Laboratory of Atomic and Subatomic Structure and Quantum Control (MOE), Guangdong Basic Research Center of Excellence for Structure and Fundamental Interactions of Matter, Institute of Quantum Matter, South China Normal University, Guangzhou 510006, China}
\affiliation{Guangdong-Hong Kong Joint Laboratory of Quantum Matter, Guangdong Provincial Key Laboratory of Nuclear Science, Southern Nuclear Science Computing Center, South China Normal University, Guangzhou 510006, China}

\author{Wen Qin \footnote{Corresponding author}} \email{qinwen@hunnu.edu.cn}
\affiliation{  Department of Physics, Hunan Normal University, and Key Laboratory of Low-Dimensional
Quantum Structures and Quantum Control of Ministry of Education,   Changsha 410081, China }
\affiliation{Synergetic Innovation Center for Quantum Effects and Applications (SICQEA), Hunan Normal University, Changsha 410081, China }

\author{Xiangyu Jiang}
\affiliation{ CAS Key Laboratory of Theoretical Physics, Institute of Theoretical Physics, Chinese Academy of Sciences, Beijing 100190, People's Republic of China }

\author{Yibo Yang}
\affiliation{School of Physical Sciences, University of Chinese Academy of Sciences,  Beijing 100049, China}
\affiliation{CAS Key Laboratory of Theoretical Physics, Institute of Theoretical Physics, Chinese Academy of Sciences, Beijing 100190, China}
\affiliation{International Centre for Theoretical Physics Asia-Pacific, Beijing/Hangzhou, 100190, China}
\affiliation{School of Fundamental Physics and Mathematical Sciences, Hangzhou Institute for Advanced Study, UCAS, Hangzhou 310024, China}

\begin{abstract}
We perform the first lattice QCD study on the radiative decay of the scalar glueball to the vector meson $\phi$ in the quenched approximation. The calculations are carried out on three gauge ensembles with different lattice spacings, which enable us to do the continuum extrapolation. We first revisit the radiative $J/\psi$ decay into the scalar glueball $G$ and obtain the partial decay width $\Gamma(J/\psi \to \gamma G)=0.578(86)~\text{keV}$ and the branching fraction $\text{Br}(J/\psi \to \gamma G) = 6.2(9)\times 10^{-3}$. We then extend the similar calculation to the process $G\to \gamma \phi$ and get the partial decay width $\Gamma(G \to \gamma \phi)= 0.074(47)~\text{keV}$, which implies that the combined branching fraction of $J/\psi\to\gamma G\to \gamma\gamma\phi$ is as small as $\mathcal{O}(10^{-9})$ such that this process is hardly detected by the BESIII experiment even with the large $J/\psi$ sample of $\mathcal{O}(10^{10})$. With the vector meson dominance model, the two-photon decay width of the scalar glueball is estimated to be $\Gamma(G\to\gamma\gamma)=0.53(46)~\text{eV}$, which results in a large stickiness $S(G)\sim \mathcal{O}(10^4)$ of the scalar glueball by assuming the stickiness of $f_2(1270)$ to be one.
\end{abstract}
\maketitle

\section{INTRODUCTION}
\label{INTRODUCTION}
Gluons and quarks are fundamental degrees of freedom of the Quantum Chromodynamics (QCD). Apart from the conventional mesons and baryons that are described by quark-antiquark ($q\bar{q}$) and three quark ($qqq$) bound states in the constituent quark models, it is usually conjectured that there also exist glueballs that are bound states of pure gluons. Glueballs are well-defined objects in the pure Yang-Mills theory, whose spectrum have been derived through the numerical lattice QCD calculations in the quenched approximation~\cite{morningstar1999glueball,chen2006glueball,Athenodorou:2020ani}. For example, the lowest lying scalar ($0^{++}$), tensor ($2^{++}$) and pseudoscalar ($0^{-+}$) glueball masses are predicted to be $1.5-1.7$~\text{GeV}, $2.2-2.4$~\text{GeV} and $2.4-2.6$~\text{GeV}, respectively. These results are supported to some extent by recent dynamical lattice QCD~\cite{Richards:2010ck,gregory2012towards,bali2000static,sun2018glueball,chen2023glueballs}. Obviously, these lowest lying glueballs share the same quantum numbers with the conventional $q\bar{q}$ mesons, and can mix with $q\bar{q}$ mesons when the gluon-quark transition is switch on. The key question is to single out the (predominant) glueball states among mesons of the same quantum numbers and similar masses.

In the scalar channel, the three $I=0$ scalar mesons, $f_0(1370)$, $f_0(1500)$ and $f_0(1710)$ are in the scalar glueball mass region. According to the SU(3) flavor symmetry, there should be only two isoscalars in a $q\bar{q}$ nonet, a surplus state hints at an additional degree of freedom that can be the lowest scalar glueball. There are many phenomenological studies on the possible mixing effects between the pure scalar glueball $G$, $s\bar{s}$ component and $n\bar{n}$ component~\cite{ochs2013status,mathieu2009physics,crede2009experimental,Giacosa:2005zt,Ren:2023ebq}. Based on their decay properties and different theoretical assumptions, either $f_0(1500)$~\cite{amsler1995evidence,close2000mixing,He:2006ij,Guo:2020akt,Close:2005vf} or $f_0(1710)$~\cite{cheng2015revisiting,Sexton:1995kd,Cheng:2006hu,Chanowitz:2005du,Llanes-Estrada:2021evz,Albaladejo:2008qa,Chen:2019buw} is assigned to be predominantly a glueball state. However, more experimental and theoretical information is desired for the scalar glueball state to be unambiguously identified.

Great efforts have been made to find the signature of glueballs in experiment. The gluon rich $J/\psi$ radiative decay is usually thought of an ideal hunting ground for glueballs. It is observed that $f_0(1710)$ is produced more copiously than $f_0(1500)$~\cite{particle2022review}. BESII and BESIII have performed partial wave analysis of the radiative decay processes $J/\psi\to \gamma X\to \gamma \pi\pi$~\cite{Ablikim:2006db}, $\gamma \eta\eta$~\cite{BESIII:2013qqz}, $\gamma K_S K_S$~\cite{BESIII:2018ubj}, and find that the yield of $f_0(1710)$ is almost one order of magnitude larger than that of $f_0(1500)$ in each individual process above. After summing over the measured branching fractions collected by PDG~\cite{particle2022review}, one has
$J/\psi\to\gamma f_0(1710))>2.1\times 10^{-3}$ and $J/\psi\to \gamma f_0(1500))>1.9\times 10^{-4}$, which can be compared with the theoretical predictions $J/\psi\to\gamma G=3.8(9)\times 10^{-3}$ from lattice QCD~\cite{gui2013scalar} and $\sim 3\times 10^{-3}$ from QCD sum rules~\cite{Narison:1996fm}. These observations support $f_0(1710)$ as a candidate for the scalar glueball. Additional evidences for this can be found in the analysis of the flavor structure in production processes of $f_0(1710)$ and $f_0(1500)$. BaBar analyzes the $\eta_c$ strong decay to three pseudoscalars and observe the enhanced $\eta'f_0(1710)$ mode and $\eta f_0(1500)$ mode\cite{BaBar:2021fkz}, BESIII observe clear $f_0(1500)$ signals but does not see significant $f_0(1710)$ signals in the $\eta\eta'$ system of the decay process $J/\psi\to\gamma\eta\eta'$~\cite{BESIII:2022iwi}. These observations indicate $f_0(1710)$ and $f_0(1500)$ are mainly flavor singlet and octet, respectively, since $\eta'(\eta)$ is mainly flavor singlet (octet) and $\eta\eta'$ only appears as a flavor octet. 


Apart from its production rate in radiative $J/\psi$ decays, the radiative decay width of the scalar glueball also provides ancillary information for the experimental search for it. However, the present theoretical results for this kind of processes is sparse and controversial. A phenomenological study based on the vector meson dominance (VMD) model gives a large partial decay width of $454~\text{keV}$~\cite{Cotanch:2004py} for $G\to \gamma\phi$, while a recent study obtains as much smaller value $14.1-29.4~\text{keV}$ using the Witten-Sakai-Sugimoto model~\cite{hechenbergerRadiativeMesonGlueball2023}. The striking discrepancy makes these predictions little informative. For example, BESIII recently performed a partial wave analysis of the $J/\psi\to\gamma\gamma \phi$ process~\cite{BESIII:2024ein} using its large ensemble of $\mathcal{O}(10^{10})$ $J/\psi$ events~\cite{BESIII:2021cxx}. Assuming an $\mathcal{O}(100~\text{MeV})$ width of the scalar glueball, the combined branching fraction $\mathrm{Br}(J/\psi\to\gamma G,G\to\gamma \phi)$ is estimated to be either $\mathcal{O}(10^{-5})$ or $\mathcal{O}(10^{-7})$ using the partial widths predicted above and the lattice prediction of $\mathrm{Br}(J/\psi\to\gamma G)$, such that no sound conclusion can be drawn.  

In this work, the process $G\to \gamma\phi$ will be investigated from lattice QCD in the quenched approximation. As the first step, we will revisit the decay process $J/\psi\to\gamma G$ following the strategy in Ref.~\cite{gui2013scalar} and compare with the previous lattice QCD result for a cross check. After that, we will extend the similar calculation to the process $G\to\gamma\phi$ to predict the partial decay width, which is expected to be less model dependent. In addition, this partial decay width can be also used to estimate the two-photon decay width $\Gamma(G\to \gamma\gamma)$ of the scalar glueball by the help of the VMD model, from which the stickiness of the scalar glueball~\cite{Chanowitz:1984cb,Crede:2008vw} can be also estimated. The practical calculation will be carried out on several large gauge ensembles different lattice spacings, which enable us to gauge the finite lattice spacing artifacts.

This work is organized as follows: Sect.~\ref{FORMALISM} introduces the formalism for calculating glueball radiative decays using the multipole expansion method. Sect.~\ref{NUMERICAL DETAILS} provides the details of the simulation on lattice QCD, including the calculations and results analysis of two-point and three-point functions. We give some discussion and conclusion in Sect.~\ref{DISCUSSION} and Sect.~\ref{SUMMARY}.

\section{FORMALISM}
\label{FORMALISM}
\begin{table*}[t]
\caption{The configuration parameters and mass spectrum. The spatial lattice spacing $a_s$ is determined from $r_{0}^{-1}=0.410(20)~\text{GeV}$ by calculating the static potential.}
\begin{ruledtabular}
\begin{tabular}{cccccc|cccc}
    \textrm{$\beta$}&
    \textrm{$\xi$}&
    \textrm{$a_s(\mathrm{fm})$}&
    \textrm{$L a_s(\mathrm{fm})$}&
    \textrm{$L^3 \times T$}&
    \textrm{$N_{\mathrm{conf}}$}&
    $m[\eta_s(0^{-+})]~\text{(GeV)}$& $m[\phi(1^{--})]~\text{(GeV)}$ & $m[f_{0}^{(s)}(0^{++})]~\text{(GeV)}$& $m[G(0^{++})]~\text{(GeV)}$\\
    \hline
    $2.4$ & 5 & $0.222(2)$ & $2.66$ & $12^3 \times 192$ & 4000 &0.7025(19)&1.0241(17)&1.569(22)&1.372(27)\\
    $2.8$ & 5 & $0.138(1)$ & $2.21$ & $16^3 \times 192$ & 4000 &0.7064(12)&1.0287(20)&1.549(29)&1.495(54)\\
    $3.0$ & 5 & $0.110(1)$ & $1.76$ & $16^3 \times 192$ & 4000 &0.6946(27)&1.0214(22)&1.593(24)&1.612(63)\\\hline
    $\infty$ &&            &        &                   &      &0.7044(20)&1.0252(23)&1.582(28) &1.635(62)
\end{tabular}
\end{ruledtabular}
\label{Tab:input parameter}
\end{table*}
In this study, we adopt the quenched lattice QCD framework to revisit the radiative decay process of $J/\psi$ to the scalar glueball $G$, namely, $J/\psi\to\gamma G$~\cite{gui2013scalar}, and then explore the possible rare decay property of the scalar glueball to the vector meson $\phi$, namely, $G\to\gamma\phi$. Both processes involve the electromagnetic (EM) transition matrix element $\langle S|J^\mu_\mathrm{em}|V\rangle$ (or its complex conjugation) between a vector ($V$) and a scalar ($S$) state, where $J_\mathrm{em}^\mu$ is the local EM current of the involved quarks (the strange quark or charm quark in this study). The explicit EM multipole expansion of this kind of matrix element reads~\cite{Dudek:2006ej}
\begin{widetext}
\begin{equation}
\begin{aligned}
    \left\langle S\left(\vec{p}_S\right)\left|J_{\mathrm{em}}^\mu(0)\right| V \left(\vec{p}_V, \lambda\right)\right\rangle
    =&\Omega^{-1}\left(Q^2\right)\left(E_1\left(Q^2\right)\left[\Omega\left(Q^2\right)   \epsilon^\mu\left(\vec{p}_V, \lambda\right)-\epsilon\left(\vec{p}_V, \lambda\right) \cdot p_S\left(p_V^\mu p_V \cdot p_S-m_{V}^2 p_S^\mu\right)\right]\right.\\
    &\left.+\frac{C_1\left(Q^2\right)}{\sqrt{-Q^2}} m_V\epsilon\left(\vec{p}_V, \lambda\right) \cdot p_S\left[p_V \cdot p_S\left(p_V+p_S\right)^\mu-m_S^2 p_V^\mu-m_{V}^2 p_S^\mu\right]\right),
\end{aligned}
\label{Eq:matrix element}
\end{equation}
\end{widetext}
where $\lambda$ refers to the polarization of the vector $V$, $Q^2$ is the squared four-momentum transfer $Q^2 \equiv -q^2 = -\left(p_V-p_S\right)^2$, and $\Omega\left(Q^2\right)=\left(p_V \cdot p_S\right)^2-m_V^2m_S^2$. There are two form factors $E_1(Q^2)$ and $C_1(Q^2)$ in the multipole decomposition but only $E_1(Q^2=0)$ enters the expression of the partial decay widths ($C_1(Q^2)$ is related to the longitudinal polarization of the photon that is unphysical)
\begin{eqnarray}
    \Gamma_{J/\psi\to \gamma G} &=& \frac{4}{27} \alpha \frac{\left|\vec{q}_{\psi\to G}\right|}{m_{\psi}^2}\left|E_1(0)\right|^2,\nonumber\\
    \Gamma_{G \rightarrow \gamma\phi}&=&\frac{1}{9} \alpha \frac{\left|\vec{q}_{G\to\phi}\right|}{m_{G}^2}\left|E_1(0)\right|^2.
\label{Eq:decay width}
\end{eqnarray}
where $\alpha$ is the fine structure constant in QED, $|\vec{q}|=\frac{m_\psi^2-m_{G}^2}{2m_\psi}$ is the magnitude of the final state photon momentum in the process $J/\psi\to\gamma G$, and $|\vec{q}|=\frac{m_G^2-m_{\phi}^2}{2m_G}$ is that for the process $G\to\gamma\phi$. The prefactors in Eq.~(\ref{Eq:decay width}) incorporate the electric charges of quarks $Q_c^2=4/9$ and $Q_s^2=1/9$, since the EM current takes the forms $J_\mathrm{em}^\mu=\bar{c}\gamma^\mu c$ and $\bar{s}\gamma^\mu s$ for $J/\psi\to \gamma G$ and $G\to \gamma \phi$, respectively. 

The matrix elements on the left hand side of Eq.~(\ref{Eq:matrix element}) can be extracted from the corresponding three-point correlation functions. Taking $G\to\gamma \phi$ for instance, we calculate the the three-point function 
\begin{equation}
\begin{aligned}
    &\Gamma_{G \rightarrow \gamma \phi}^{(3), \mu i}\left(\vec{p}_i, t_i=0 ; \vec{p}_f, t_f ; \vec{q}, t\right) \\
    &=\sum_{\vec{x}, \vec{y}} e^{-i \vec{p}_f \cdot \vec{y}} e^{i \vec{p}_i \cdot \vec{x}} \left\langle\mathcal{O}^i_{\phi}\left(\vec{y}, t_f\right) J_{\mathrm{em}}^\mu(\vec{0}, t) \mathcal{O}_{G}^{\dagger}\left(\vec{x}, t_i\right)\right\rangle,
\end{aligned}
\label{Eq:Corr3_phi}
\end{equation}
where $\mathcal{O}_{G}$ and $\mathcal{O}^i_{\phi}$ are the interpolation field operator of scalar glueballs $G$ and vector mesons. By inserting complete sets of states, the three-point function can be related to matrix elements as
\begin{equation}
\begin{aligned}
    \Gamma_{G \rightarrow \gamma\phi}^{(3), \mu i}&\left(\vec{p}_{i}, t_{i}=0 ; \vec{p}_{f}, t_{f} ; \vec{q}, t\right) \\
    =\sum_{m, n, \lambda_{m}} &\frac{e^{-E_{m} (t_{f}-t)} e^{-E_{n} t}}{4 V E_{m} E_{n}} \left\langle\Omega\left|\mathcal{O}^{i}_\phi(0)\right| m, \vec{p}_{f},\lambda_m\right\rangle \\
    &\times\left\langle m, \vec{p}_{f},\lambda_m\left|J^{\mu}_{\mathrm{em}}(0)\right| n, \vec{p}_{i} \right\rangle\left\langle n, \vec{p}_{i}\left|\mathcal{O}_{G}^{\dagger}(0)\right| \Omega\right\rangle \\
    \stackrel{t_{f} \gg t>0}{\longrightarrow} &\frac{e^{-E_{\phi} (t_f-t)} e^{-E_{G} t}}{4 V E_{\phi} E_{G}}  Z^{(\phi),i}Z^{(G)} \\
    &\times\left\langle \phi\left(\vec{p}_{f},\lambda_{\phi}\right)\left|J_{\mathrm{em}}^{\mu}(0)\right|G\left(\vec{p}_{i}\right)\right\rangle,
\label{Eq:Parameterization of three-point function}
\end{aligned}
\end{equation}
where $E_G$ and $E_{\phi}$ are the energy of ground scalar glueball $G$ and $\phi$ meson, respectively, and the overlap factors $Z^{(\phi)}(\vec{p}_f) = \left\langle \Omega \left| \mathcal{O}^i_f(0)\right|\phi\left(\vec{p}_{f},\lambda\right)\right\rangle$ and $Z^{(G)}(p_i) = \left\langle G\left(\vec{p}_{i})\right) \left|\mathcal{O}_{G}^{\dagger}(0)\right| \Omega\right\rangle$ can be obtained by fitting the corresponding two-point functions. For example, the $\phi$ meson two-point function is like
\begin{eqnarray}
    \Gamma^{(2), ij}(\vec{p},t)&=&\sum_{\vec{x}} e^{-i \vec{p} \cdot \vec{x}}\left\langle\Omega\left|\mathcal{O}^i_\phi(\vec{x}, t) \mathcal{O}^{j,\dagger}_{\phi}(0,0)\right| \Omega\right\rangle\nonumber\\
    &\stackrel{t \rightarrow \infty}{\longrightarrow}& \frac{|Z^{(\phi)}|^2}{2E(\vec{p})}e^{-E(\vec{p})t}\sum\limits_{\lambda_\phi}\epsilon^{j*}(\vec{p},\lambda_\phi)\epsilon^{i}(\vec{p},\lambda_\phi)\nonumber\\
    &=& \frac{|Z^{(\phi)}|^2}{2E(\vec{p})}e^{-E(\vec{p})t}\left(\delta^{ij}+\frac{p^ip^j}{m_\phi^2}\right).
    \label{Eq:two point function of phi}
\end{eqnarray}

Therefore, by directly calculating the corresponding three-point and two-point functions on the lattice, the transition matrix element can be obtained from Eq.~(\ref{Eq:Parameterization of three-point function}), and the form factors at different $Q^2$ can be solved based on the multipole decomposition formula in Eq.~(\ref{Eq:matrix element}), and finally, the on-shell form factor $E(0)$ can be obtained through the interpolation or extrapolation to $Q^2=0$, from which the decay width can be predicted. To control the discretization error, we will perform calculations on three different lattice spacings and extrapolate the form factors to the continuum limit.

\section{NUMERICAL DETAILS}
\label{NUMERICAL DETAILS}
\subsection{Lattice setup}
We performed simulation in the quenched approximation lattice QCD. Three ensembles with different gauge couplings $\beta$ generated using the anisotropic tadpole improved Symanzik’s gauge action~\cite{Morningstar:1997ff,Morningstar:1999rf,Chen:2005mg}. Each ensemble has 4000 configurations to get a good statistical signal. The bare anisotropy $\xi = a_s / a_t$ is set to 5 so that there is a better resolution in the time direction. The spatial lattice spacing is obtained from the static quark-antiquark potential. All configuration parameters are listed in Tab.~\ref{Tab:input parameter}. The quark propagators were computed using anisotropic clover fermion action~\cite{Zhang:2001in,Su:2004sc,CLQCD:2009nvn}. The tadpole improved tree-level value is used for the clover coefficient $c_{sw}$ and the bare velocity of light $\nu_s $ have been tuned by the vector meson dispersion relation. We tuned the bare strange quark mass parameter on each ensemble to give the physical mass of $\phi$, $m_\phi = 1.02~\text{GeV}$. Using these quark parameters, we calculated the spectra of strangeonium, including pseudoscalar ($J^{PC}=0^{-+}$) meson $\eta_s$, vector ($1^{--}$) meson $\phi$ and the scalar ($0^{++}$) mesons across the three ensembles, which are also listed in Tab.~\ref{Tab:input parameter}. The bare charm quark masses for the three lattices are set by the physical mass of $J/\psi$, $m_{J/\psi}=3.097~\text{GeV}$. 

\subsection{Two-point function}

\begin{figure}[t!]

\centering
    \includegraphics[scale=0.45]{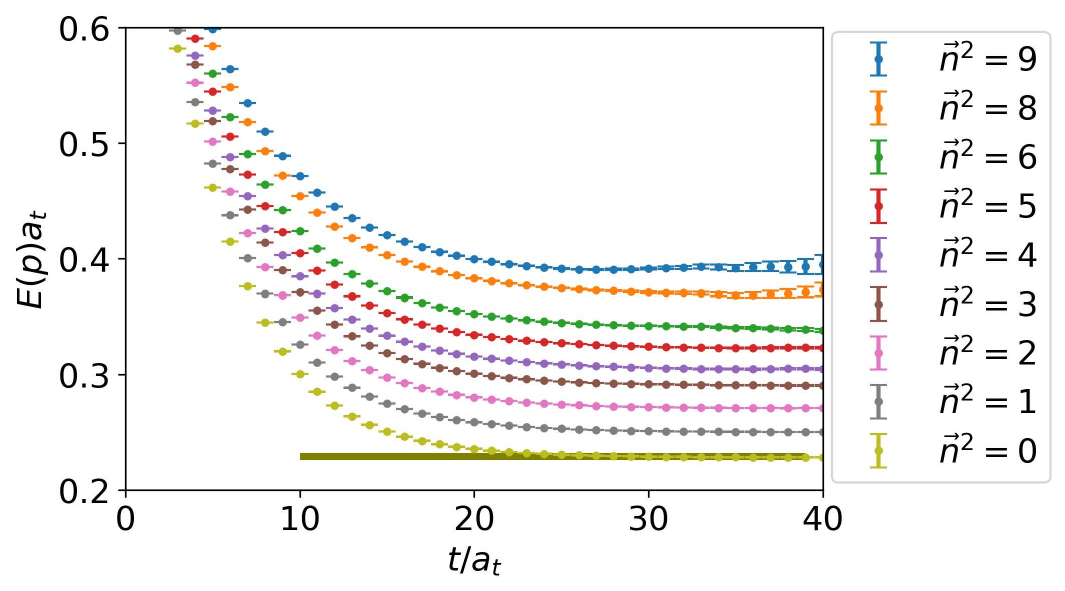}
    \includegraphics[scale=0.45]{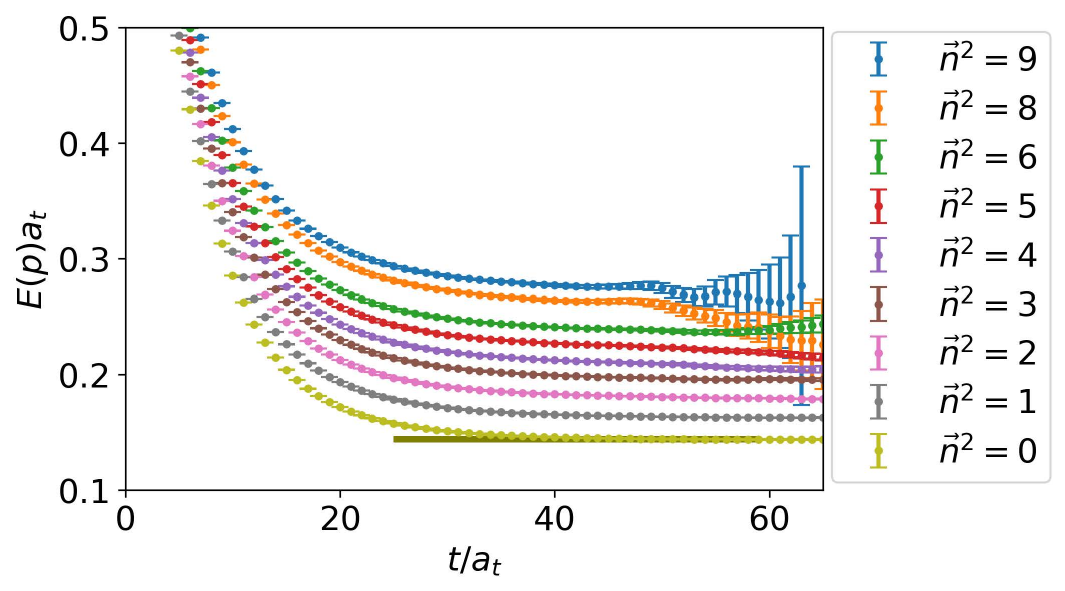}
    \includegraphics[scale=0.45]{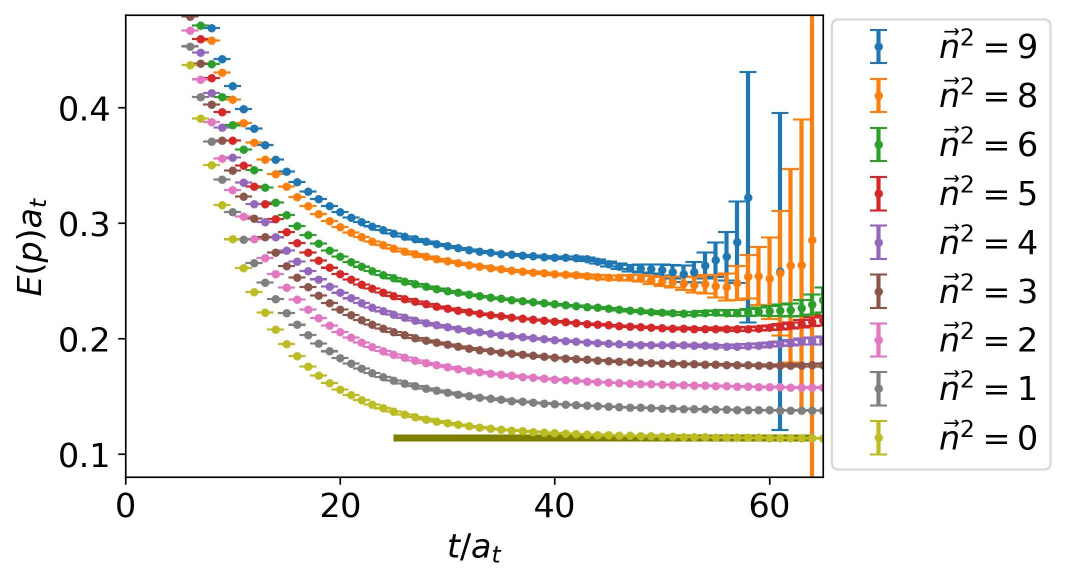}
\caption{The mass plateaus for distinct momentum modes, $\vec{p} = \frac{2 \pi \vec{n}}{L a_s}$ are shown from top to bottom, corresponding to $\beta = 2.4, 2.8, 3.0$  respectively.}\label{Fig:Effective_mass_phi}
\end{figure}

\begin{figure}[t!]
\centering
    \includegraphics[scale=0.4]{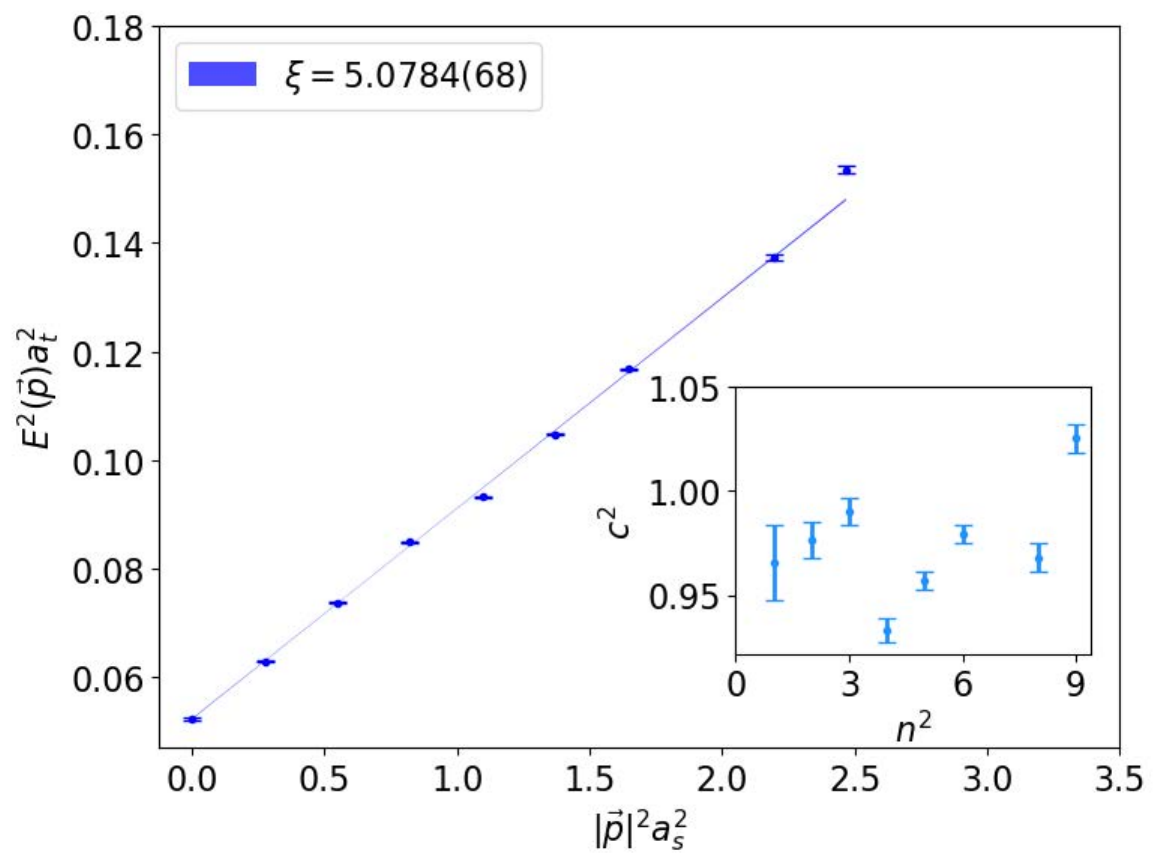}
    \includegraphics[scale=0.4]{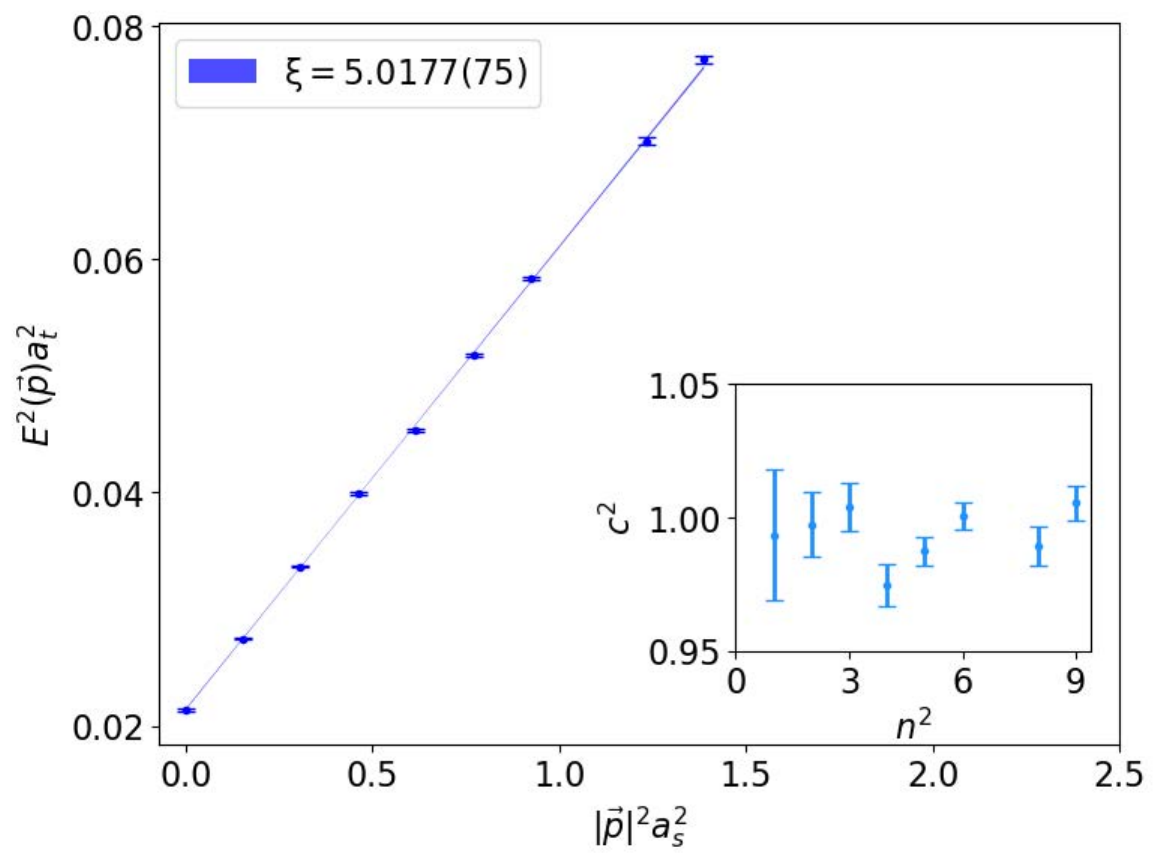}
    \includegraphics[scale=0.4]{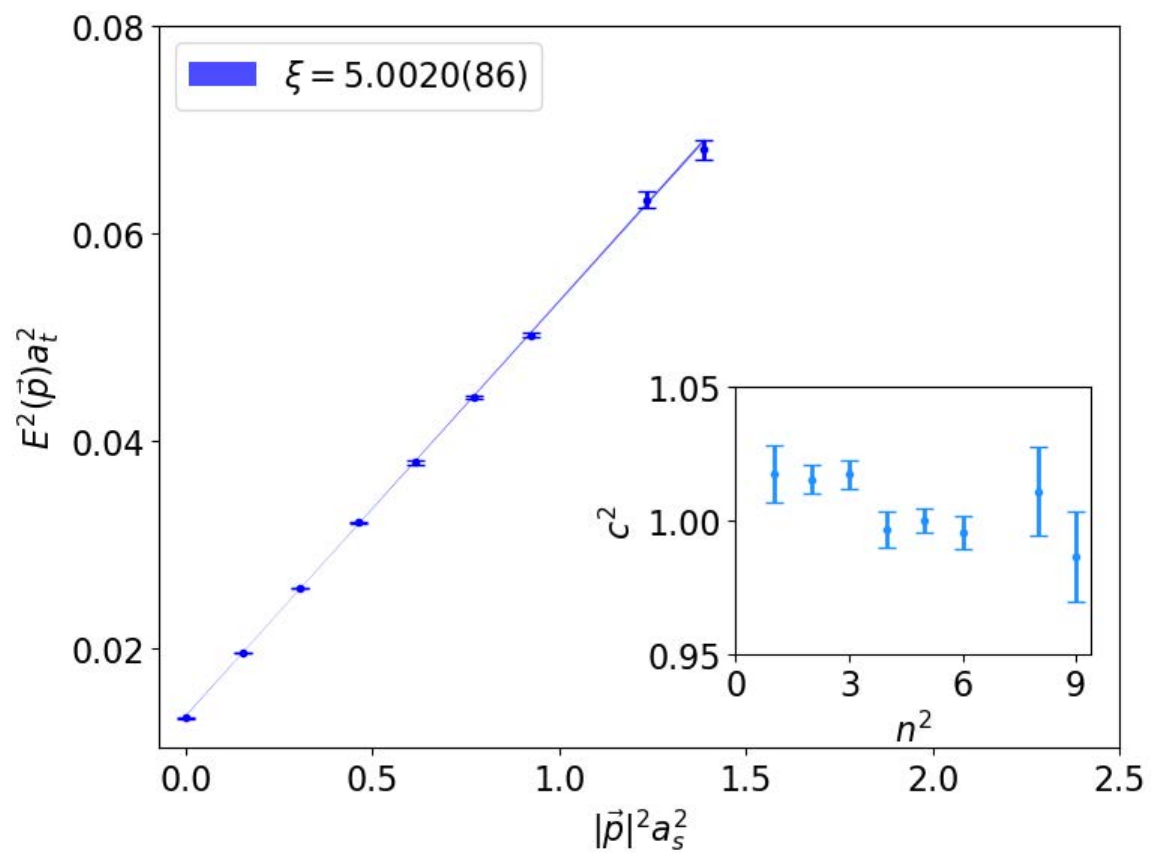}
\caption{Dispersion relationship of $\phi$ mesons on three ensembles. The results from top to bottom are for $\beta=2.4$, $\beta=2.8$ and $\beta=3.0$. Two methods are used to verify the dispersion relationship. When assuming the speed of light is equal to 1 and fitting the dispersion relationship by $E_\phi^2(\vec{p})a_t^2=m_\phi^2a_t^2+\frac{1}{\xi^2}|\vec{p}|^2a_s^2$, the resulting anisotropy $\xi$ values are 5.08, 5.02, and 5.00, respectively, as shown in the larger graphs. Alternatively, using a bare anisotropy of $\xi = 5$, the speed of light is calculated by $c^2=\frac{E^2(\vec{p})-m_{\phi}^2}{|\vec{p}|^2}$, as depicted in the smaller graphs.}
\label{Fig:dispersion relation}
\end{figure}

\begin{figure}[t!]
\centering
    \includegraphics[scale=0.4]{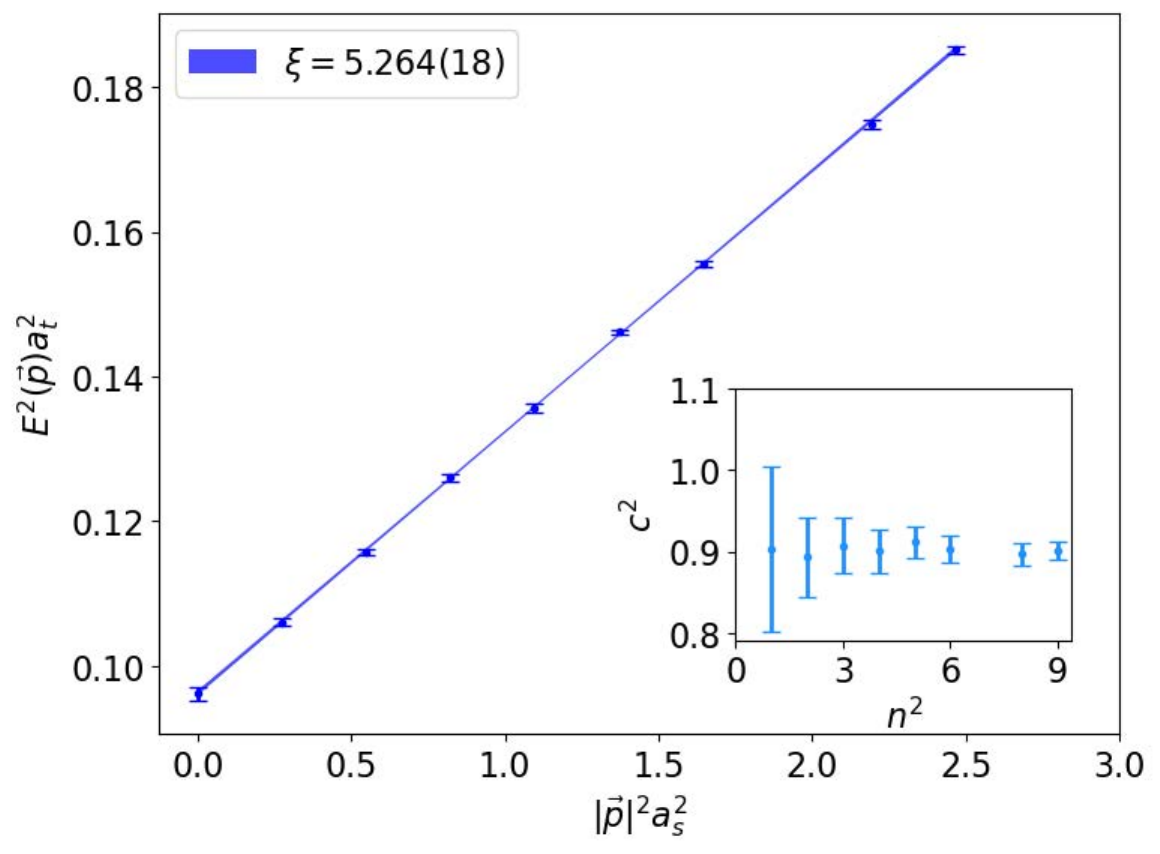}
    \includegraphics[scale=0.4]{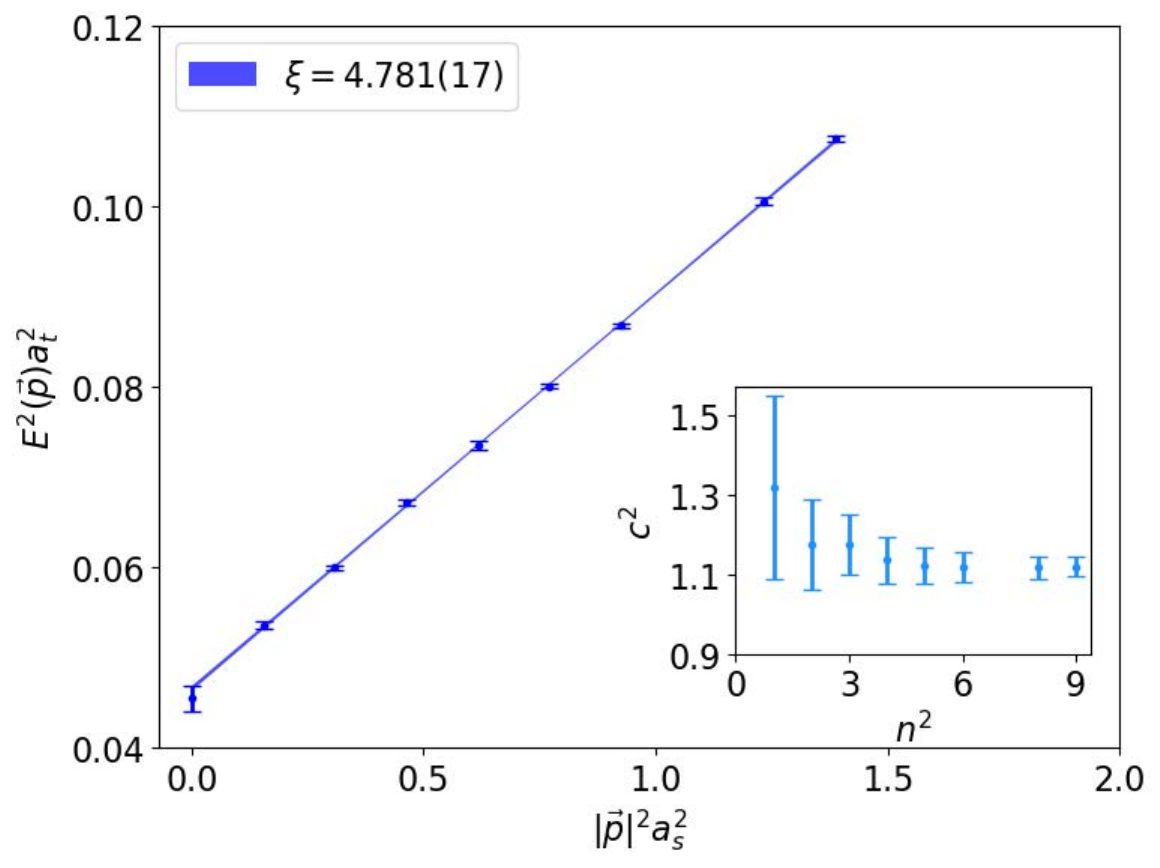}
    \includegraphics[scale=0.4]{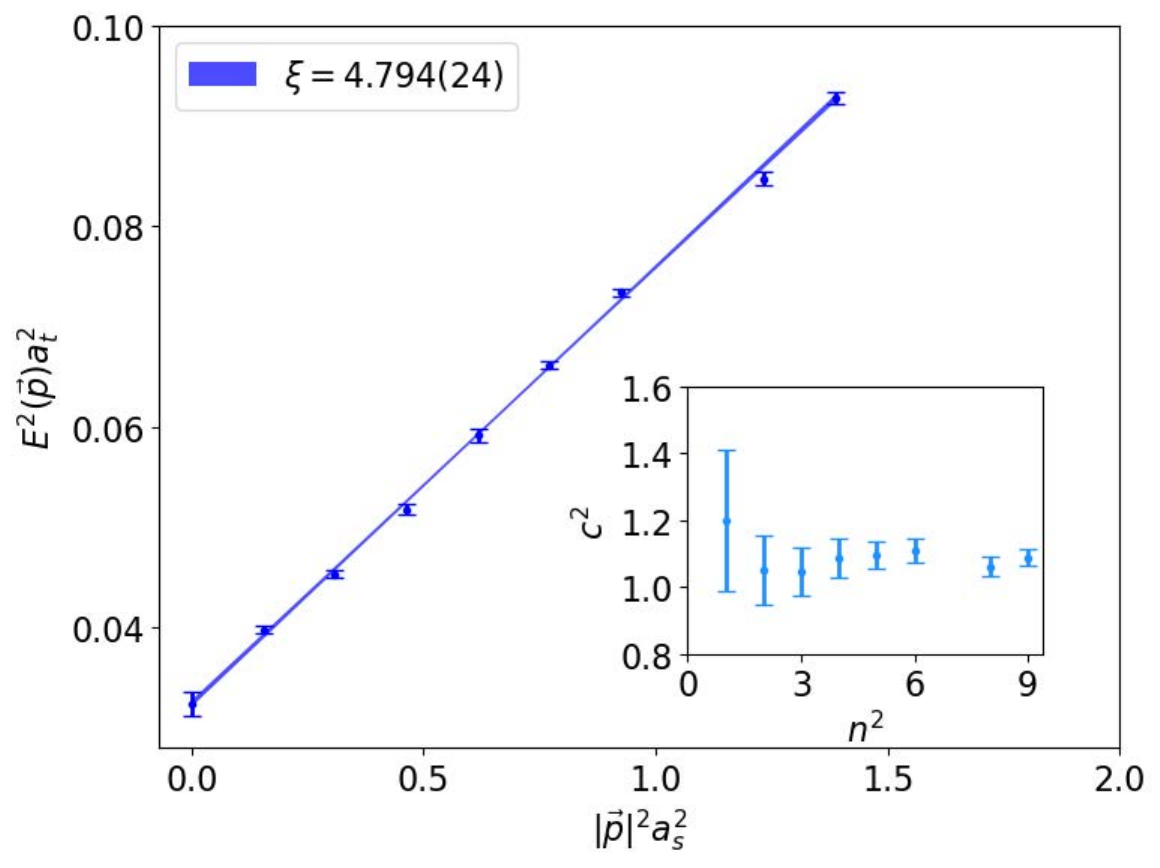}
\caption{Dispersion relationship of glueball on three ensembles. The results from top to bottom are for $\beta=2.4$, $\beta=2.8$ and $\beta=3.0$.}
\label{Fig:dispersion relation glueball}
\end{figure}
In order to determine the masses of the $\phi$ meson and glueball, it is necessary to calculate the corresponding two-point function. The local fermion bilinear operators, denoted as $\mathcal{O}=\bar{\psi} \Gamma \psi$, are utilized to compute the mass spectra of strangeonium, as outlined in Tab.~\ref{Tab:input parameter}. To enhance the signal-to-noise ratio, point sources are placed on each time slice to calculate propagators, and subsequently, the resulting two-point functions are averaged. 

Two mass terms are employed to fit the two-point functions. The fit results and the effective masses of the two-point functions are illustrated in Fig.~\ref{Fig:Effective_mass_phi}, and the fit values are detailed in Tab.~\ref{Tab:input parameter}. As previously mentioned, the bare speed of light parameters $\nu_s$ are adjusted according to the meson dispersion relation. Similarly to the approach used in~\cite{Jiang:2022gnd}, the dispersion relation can also be used to determine the anisotropy parameter $\xi$. It is observed that the fitted anisotropy $\xi$ deviates by less than $2\%$ from 5 when assuming the speed of light is 1, and the speed of light is very close to 1 when setting the anisotropy $\xi = 5$. These findings are depicted in Fig.~\ref{Fig:dispersion relation}.

\begin{figure}[t!]
\centering
    \includegraphics[scale=0.45]{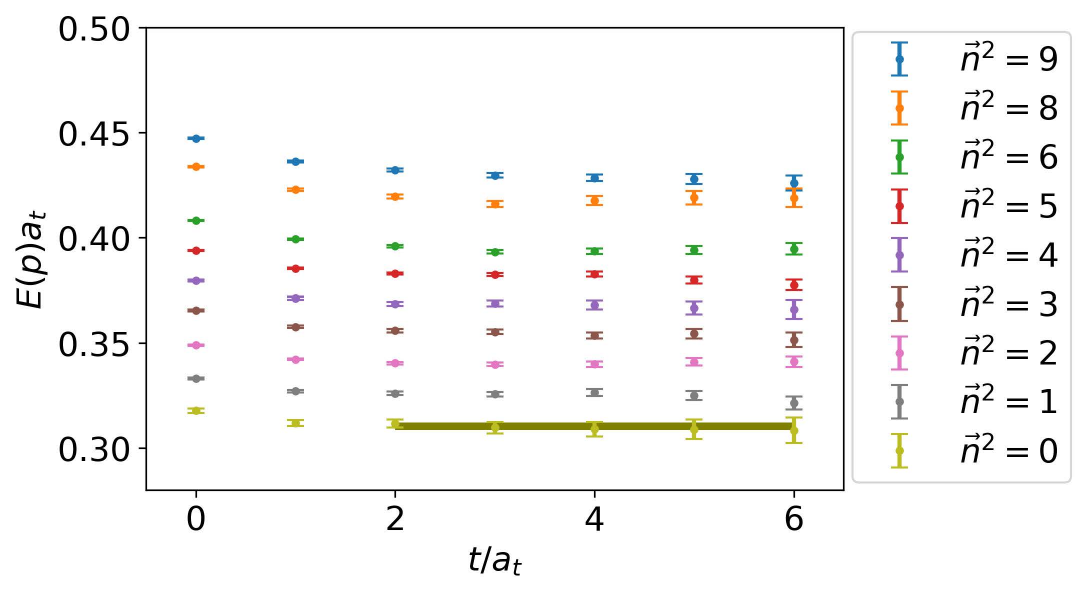}
    \includegraphics[scale=0.45]{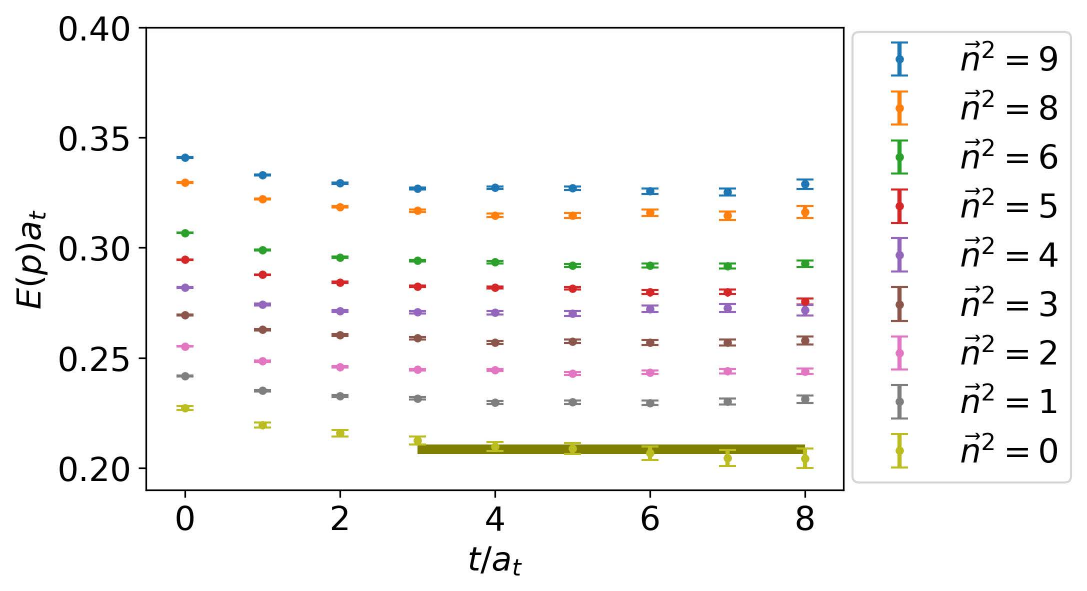}
    \includegraphics[scale=0.45]{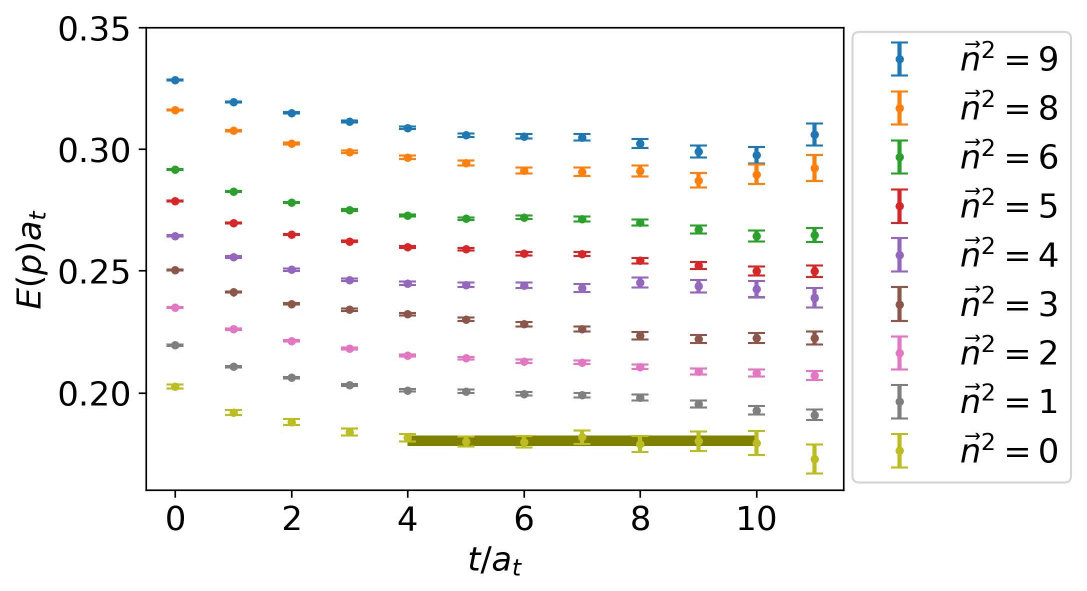}
\caption{The effective masses of the glueball two-point functions constructed using optimized operators for distinct momentum modes $\vec{p}=\frac{2\pi\vec{n}}{La_s}$ are shown from top to bottom, corresponding to $\beta=2.4$, $\beta=2.8$ and $\beta = 3.0$,respectively.}
\label{Fig:Mass glueball}
\end{figure}

We employ the method for constructing scalar glueball operators as outlined in reference~\cite{gui2013scalar}. By constructing 24 glueball operators set $\{\mathcal{O}_{G,i}, i=1,2,...,24\}$ comprised of Wilson loops satisfying the $A_{1}^{++}$ representation of the discrete spatial point group, we assemble the two-point function matrix $\Gamma_{G,ij}^{(2)}(t,\vec{p})=\langle\mathcal{O}_{G,i}(t,\vec{p})\mathcal{O}^{\dagger}_{G,j}(0)\rangle$. By performing a variational analysis on this $24\times24$ correlation function matrix and solving the generalized eigenvalue equation, we obtain optimized operators $\mathcal{O}_{G,\lambda}=\sum_{i}c_{\lambda,i}\mathcal{O}_{G,i}$ that project predominantly onto single states, where $c_{\lambda,i}$ are the eigenvectors corresponding to the eigenvalue $\lambda$ of the generalized eigenvalue equation. Given the substantial statistical fluctuations of the glueball operators, the construction of optimized glueball operators is crucial for studying the glueball spectrum and its decay properties. We performed a single mass term fit to the optimized glueball  two-point function normalized using $\Gamma_G^{(2)}(t=0,\vec{p})$, and found that the overlap factor $Z_G$ is very close to 1. This indicates that the optimized glueball operator almost projects onto the ground scalar glueball. Furthermore, given the short effective fitting range for the scalar glueball, we also considered the systematic errors arising from different choices of fitting intervals. The dispersion relationship and effective mass results from the optimized glueball operators are shown in Fig.~\ref{Fig:dispersion relation glueball} and Fig.~\ref{Fig:Mass glueball} 
respectively, and the glueball mass obtained from our fitting is presented in Tab.~\ref{Tab:input parameter}. These optimized glueball operators will be further utilized for constructing the three-point functions discussed later.

\subsection{$J/\psi \to \gamma G$}
\begin{figure}[t]
\centering
    \includegraphics[scale=0.48]{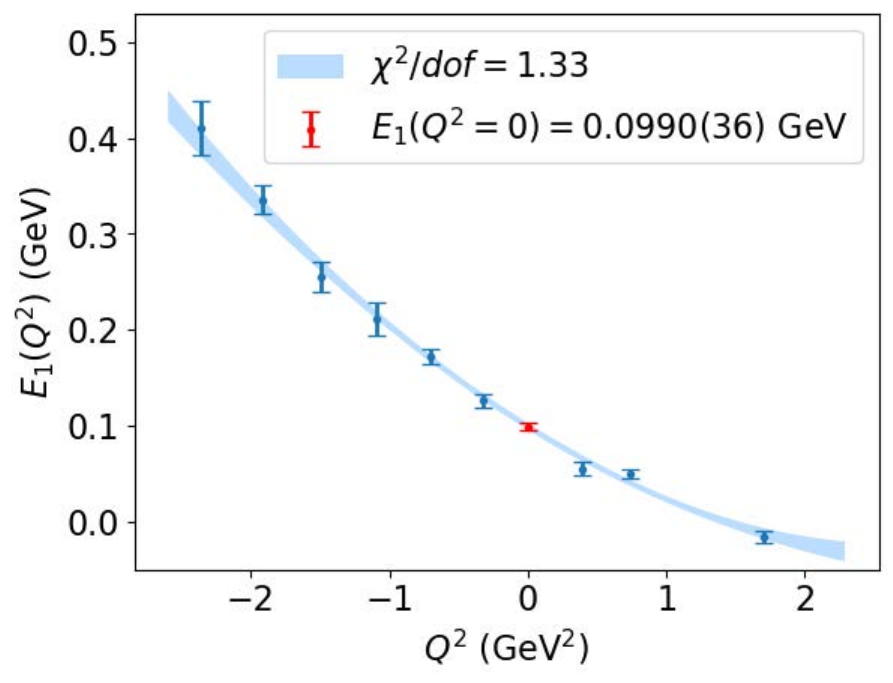}
    \includegraphics[scale=0.48]{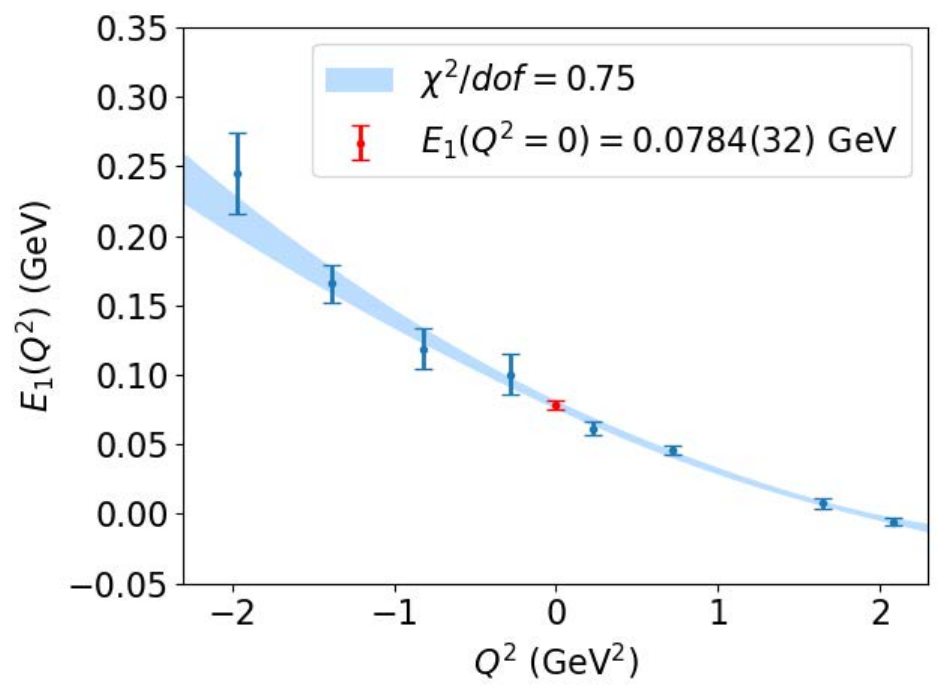}
    \includegraphics[scale=0.48]{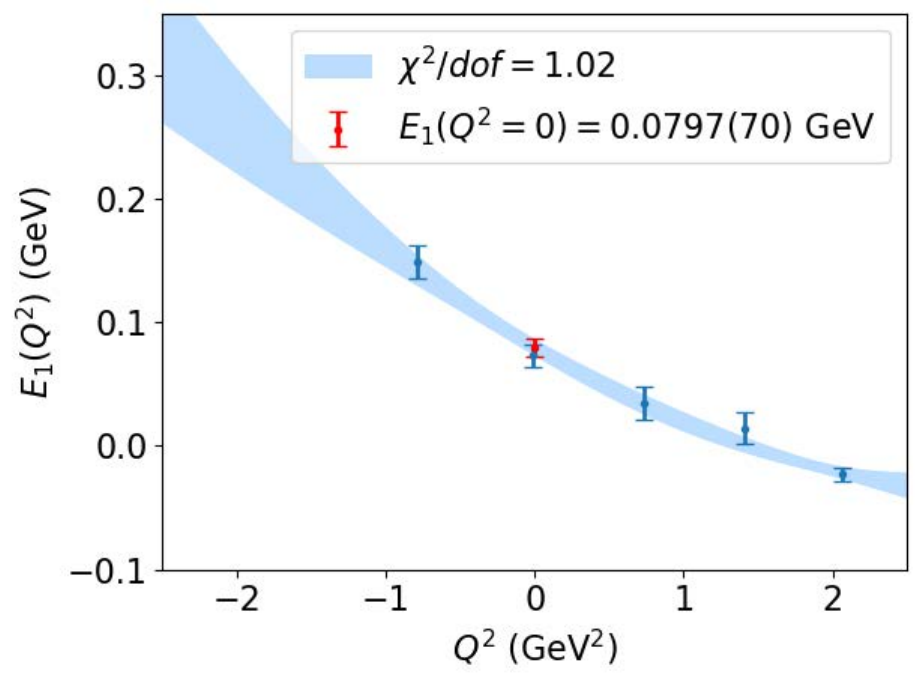}
\caption{Extrapolation of $E_1(Q^2)$ to the on-shell point $Q^2 = 0$ for the $J/\psi \to \gamma G$ process. The results from left to right correspond to $\beta=2.4$, $\beta=2.8$ and $\beta=3.0$.}
\label{Fig:form factors of jpsi to scalar glueball}
\end{figure}
Before calculating $G\to \gamma \phi$, as a check, we first computed the process $J/\psi \to \gamma G$. For this purpose, we calculate the following three-point function:
\begin{eqnarray}
    &&\Gamma_{J/\psi\to\gamma G}^{(3),\mu}\left(\vec{p}_{i},0;\vec{p}_{f},t_{f};\vec{q},t\right)\nonumber\\
    & =&\langle\mathcal{O}_{G}(\vec{p}_{f},t_{f})J^{\mu}_\text{em}(\vec{q},t)\mathcal{O}_{J/\psi}^{\dagger}(\vec{p}_{i},0)\rangle.
\end{eqnarray}
Here, we place the $J/\psi$ operator $\mathcal{O}_{J/\psi}$ at $t_i=0$, the current operator at time $t$, and the optimized scalar glueball operator at time $t_f$. Due to our use of the optimized glueball operator, we were able to set $t_f$ very close to $t$. In fact, we tested three scenarios with $\Delta t\equiv t_f-t = 0, 1, 2$, and found that the results did not show a significant dependence on $\Delta t$. In subsequent calculations, we use the results with $\Delta t=2$ for $\beta=2.4$ and $\beta=2.8$, and the result with $\Delta t = 5$ for $\beta=3.0$. The difference from our previous work~\cite{gui2013scalar} is that we use a wall source when computing the quark propagator, which reduces the accessible values of $Q^2$, but provides a better signal-to-noise ratio. We define the ratio
\begin{align}
    R^{\mu i}(t,t_f-t)=&\frac{\Gamma_{J/\psi\to\gamma G}^{(3)\mu i}\left(\vec{p}_{f},\vec{q},t_{f},t\right)}{\Gamma_{J/\psi}^{(2)}(\vec{p}_{i}=\vec{q}+\vec{p}_{f},t)\Gamma_{G}^{(2)}(\vec{p}_{f},t_{f}-t)} \nonumber \\ \times &\sqrt{4E_{J/\psi}(\vec{p}_{i})E_{G}(\vec{p}_{f})W_{J/\psi}(\vec{p}_{i})W_{G}(\vec{p}_{f})}, 
\label{Eq:ratio}
\end{align}
where $\Gamma_{J/\psi}^{(2)}$, $\Gamma_G^{(2)}$ are $J/\psi$ and scalar glueball two-point correlation. The overlap factors $W_{J/\psi}(\vec{p}_i)= \frac{Z_{J/\psi}^{2}}{2E_{J/\psi}V}$, $W_{G}(\vec{p}_{f})=\frac{Z_{G}^{2}}{2E_{G}V}$, as well as the energy of $J/\psi$ $E_{J/\psi}(\vec{p}_i)$ and the scalar glueball $E_G(\vec{p}_f)$ can be obtained by fitting from corresponding the two-point functions. It should be noted that, the local vector current $J_\text{em}^\mu=\bar{c}\gamma^\mu c$ here and $J_\text{em}^\mu=\bar{s}\gamma^\mu s$ for the process $G\to\gamma\phi$, which are conserved in the continuum,  are no longer conserved on the finite lattice and requires multiplicative normalization factors $Z_V^\mu$, which can be determined following the strategy in  Refs.~\cite{Dudek:2006ej,Yang:2012mya}
\begin{equation}
Z_{V_{f=s,c}}^{(\mu)}(t) = \frac{p^{\mu}}{E(\vec{p})}\frac{\Gamma_{\eta_{f}}^{(2)}\left(\vec{p};t_{f}\right)}{\Gamma_{\eta_{f}}^{(3),\mu}\left(\vec{p},\vec{p},t_{f},t\right)}.
\end{equation}
Only the spatial components $J_\text{em}^{i=x,y,z}$ are involved in our calculation, whose renormalization factors ($Z_{V_c}^{(i)}$ for $\bar{c}\gamma_i c$ and $Z_{V_s}^{(i)}$ for $\bar{s}\gamma_i s$) are collected in Tab.~\ref{Tab:Renormalization}


\begin{table}[h]
\caption{\label{tab:table1}
The renormalization constants for the electromagnetic current  of charm and strange quark on three lattices.}
\begin{ruledtabular}
\begin{tabular}{ccccc}
\textrm{$\beta$}&
\textrm{$2.4$}&
\textrm{$2.8$}&
\textrm{$3.0$}&\\
\colrule
$Z_{V_c}^{(i)}$\footnote{The values of $Z^{(i)}_{V_c}$ listed in Refs.~\cite{Yang:2012mya,gui2013scalar,gui2019study} include an additional factor $\nu_s$ to suppress the discrepancy between $Z_{V_c}^{(i)}$ and $Z_{V_c}^{(t)}$, which would cancel part of the discretization errors, especially on two coarse lattices with $\beta=2.4$ and 2.8~\cite{Yang:2012mya,gui2013scalar}. When we included one more fine lattice $(\beta=3.0)$ to the calculation of the decay process $J/\psi\to \gamma G_{0^{-+}}$~in Ref.~\cite{gui2019study}, we realized the $\nu_s$ factor had been inappropriately added and was removed in the practical calculation. The values of $Z_{V_c}^{(i)}$ in Table~\ref{tab:table1} are the ones after removing the $\nu_s=0.71,0.74,0.77$ at $\beta=2.4,2.8,3.0$, respectively. The values from Refs. \cite{Yang:2012mya, Yang:2013xba} should be updated accordingly.} & $1.955(21)$ & $1.500(10)$ & $1.379(8)$ \\
\end{tabular}
\end{ruledtabular}
\label{Tab:Renormalization}
\end{table}

According to Eq.~(\ref{Eq:Parameterization of three-point function}) and Eq.~(\ref{Eq:two point function of phi}), we can use the multipole expansion formula Eq.~(\ref{Eq:matrix element}) to extract the form factor $E_1(Q^2,t)$ from $R^{\mu i}(t)$. We compute all rotationally equivalent momentum combinations and average all of them. To reduce the influence of $\phi$ excited states, we use the following formula  to fit $E_1(Q^2,t)$ to obtain $E_{1}(Q^{2})$:
\begin{equation}
    E_{1}(Q^{2},t) = E_{1}(Q^{2}) + A(Q^2)e^{-E t}.
\end{equation}

\begin{figure}[t!]
\centering
    \includegraphics[scale=0.48]{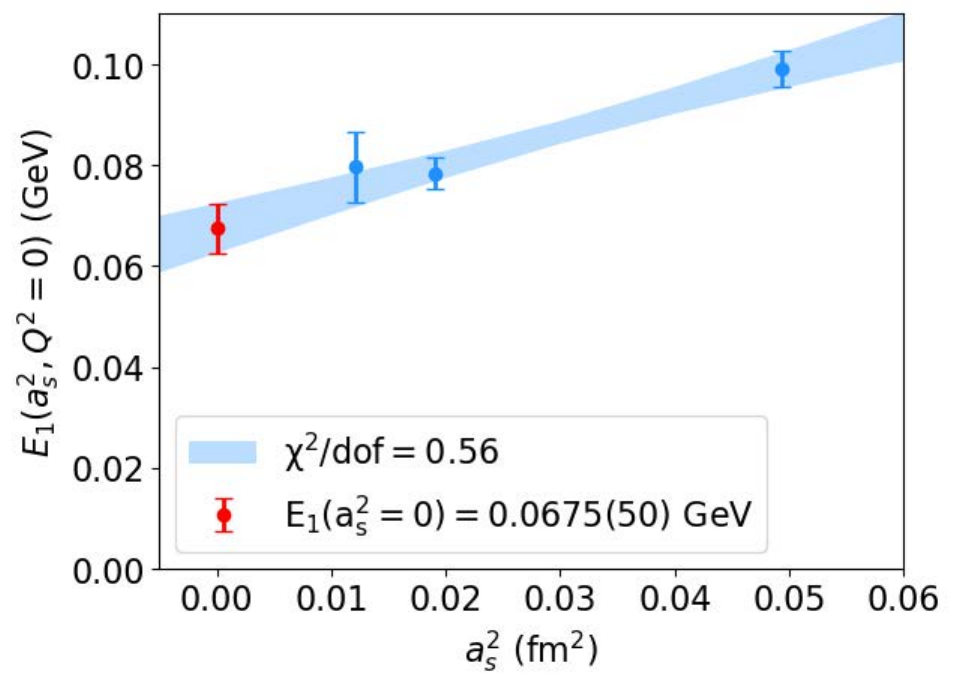}
\caption{The continuum limit extrapolation of the form factors for the $J/\psi \to \gamma G$ process is carried out at three different lattice spacings using a linear extrapolation.}
\label{Fig:Linear extrapolation of form factor of jpsi to scalar glueball}
\end{figure}
Once we obtain $E_1(Q^2)$ at different $Q^2$, we need to interpolate to the on-shell point where $Q^2 = 0 $. Here, we use a polynomial for fitting as used in Ref.~\cite{gui2013scalar},
\begin{equation}\label{eq:interpolation}
    E_{1}(Q^{2})=E_{1}(0)+a Q^{2}+b Q^{4}.
\end{equation}
Our fitting is shown in Fig.~\ref{Fig:form factors of jpsi to scalar glueball}. Considering the applicability of polynomial interpolation, we present the interpolation results for data points within $Q^2 \in [-2, 2]~\text{GeV}^2 $.
 By using wall source to compute the charm quark propagator, we effectively improve the signal of three point function, resulting in a smoother variation of the form factor $E_1(Q^2)$ with $Q^2$ compared to the literature~\cite{gui2013scalar}. After obtaining $E_1(0)$ for each ensemble, we further extrapolate the results to the continuum limit using a linear function:
\begin{equation}
    E_1(0,a) = E_1^{\text{cont.}}(0) + A a_s^2.
\end{equation}
The results on three ensembles and extrapolation are shown in Fig.~\ref{Fig:Linear extrapolation of form factor of jpsi to scalar glueball}. Finally, we obtain the continuum limit value of
\begin{equation}
    E_1(0)^{\text{cont. }}= 0.0675(50)~\text{GeV},
\end{equation}
which corresponds to the partial decay width of
\begin{equation}
    \Gamma_{J/\psi \rightarrow \gamma G} =0.578(86)~\text{keV}.
\end{equation}
Due to the change in current renormalization, our result is slightly larger than the previous result, $\Gamma_{J/\psi\to \gamma G}=0.35(8)~\text{keV}$~\cite{gui2013scalar}, but it is consistent within $2\sigma$.

\subsection{$G \to \gamma \phi$}

Based on the Eq.~\ref{Eq:Corr3_phi}, for the calculation of the process $G \to \gamma\phi$, we need to place the glueball operator at the source, that is, at the time slice $t_i=0$, and then multiply it by the two-point correlation function of the current and $\phi$ operators
\begin{align}
    \Gamma_{G\to\gamma\phi}^{(3),\mu}(t_{f},\vec{p}_{f};t,\vec{q})&=\langle\mathcal{O}_{\phi}(\vec{p}_{f},t_{f})J_{em}^{s,\mu}(\vec{q},t)\mathcal{O}_{G}^{\dagger}(\vec{p}_{i},0)\rangle \nonumber\\
	    &=\sum_{x,y,z}e^{-i\vec{p}_{f}z}e^{i\vec{q}(x+y)}\langle\bar{\psi}(z,t_{f})\gamma_{i}\psi(z,t_{f}) \nonumber\\
	    &\times\bar{\psi}(x,t)\gamma_{\mu}\psi(y,t)\mathcal{O}_{G}^{\dagger}(\vec{p}_{i},0)\rangle.
\end{align}
However, in practical calculations, we find that the three-point function obtained in this way has very large statistical noise, making it difficult to obtain an effective signal. 
we can also construct the three-point function of this process similarly to the $J/\psi \to \gamma G$ process, that is,
\begin{align}
\Gamma_{\phi\gamma\to G}^{(3),\mu}(t_{f},\vec{p}_{f};t,\vec{q})	&=\langle\mathcal{O}_{G}(\vec{p}_{f},t_{f})J_{em}^{s,\mu}(\vec{q},t)\mathcal{O}_{\phi}(\vec{p}_{i},0)\rangle \nonumber\\
	&=\sum_{x,y,z}e^{i\vec{q}z}e^{i\vec{p}_{i}(x+y)}\langle\mathcal{O}_{G}^{\dagger}(\vec{p}_{f},t_{f}) \nonumber \\
	\times \bar{\psi}&(z,t)\gamma_{\mu}\psi(z,t)\bar{\psi}(x,0)\gamma_{i}\psi(y,0)\rangle.
\end{align}
We find that the three-point correlation functions constructed in this way have a better statistical signal. Thus, the computation procedure is similar to that of $J/\psi \to \gamma G$.

\begin{figure*}[t]
    \centering
    \includegraphics[scale=0.35]{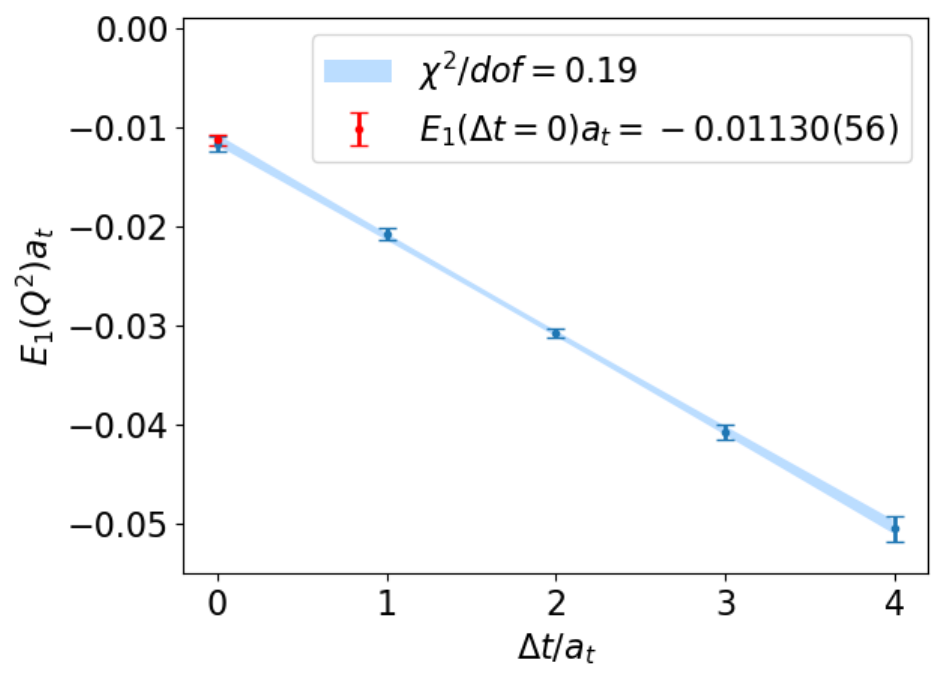}
    \includegraphics[scale=0.35]{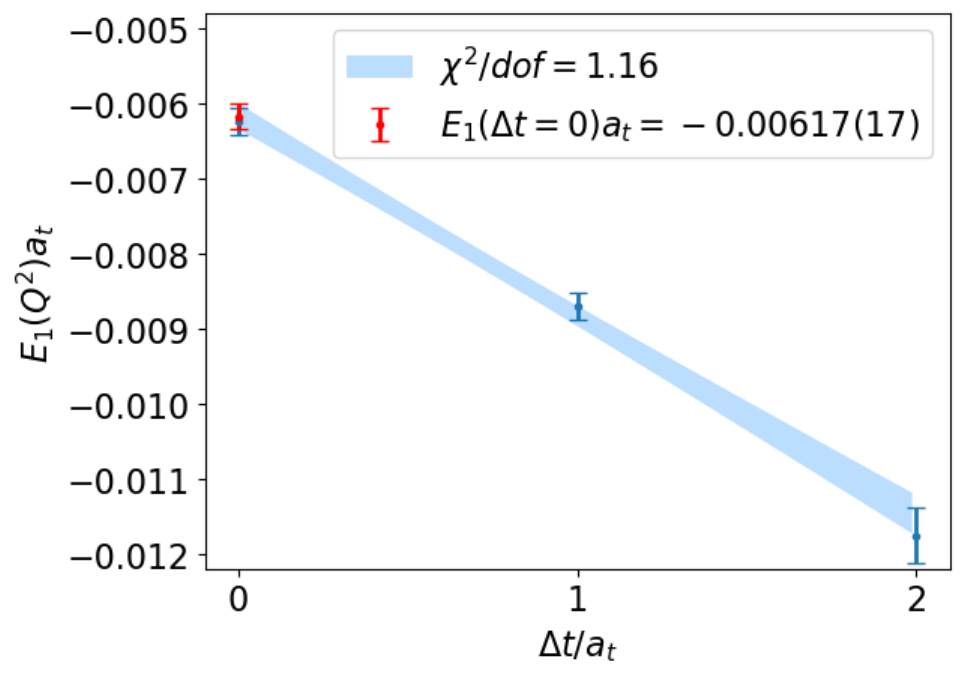}
    \includegraphics[scale=0.35]{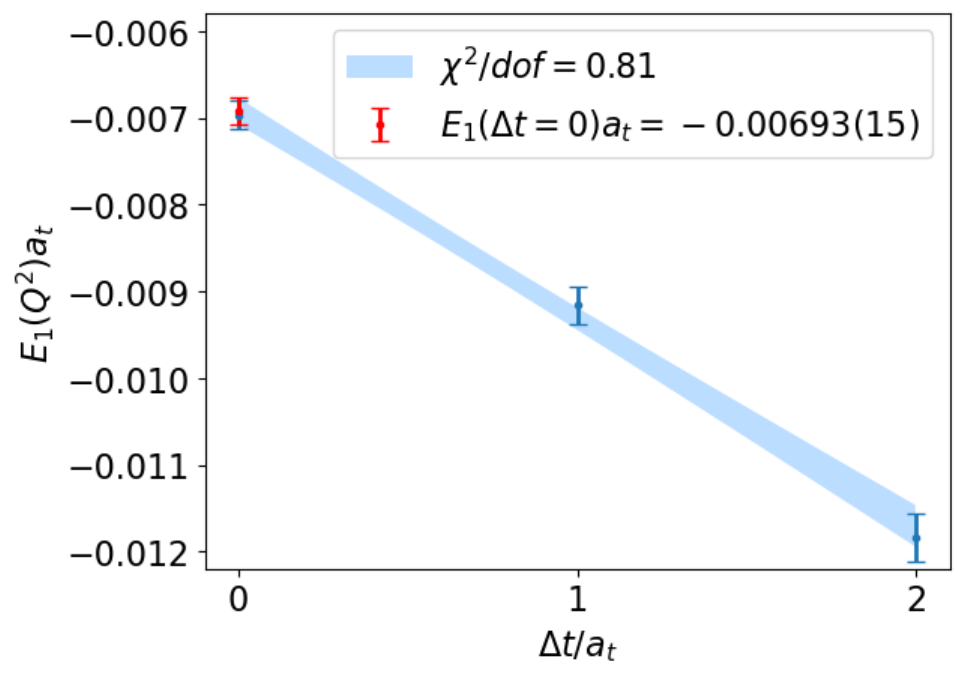}
\caption{The variation of the form factors $E_1(Q^2)$ with the time interval $\Delta t =t_f -t$ is presented. The cases for $\beta=2.4$, $\beta=2.8$ and $\beta=3.0$ are shown sequentially from left to right. For the cases of $\beta=2.8$ and $\beta=3.0$, only three time intervals $\Delta t=0,1,2$ were investigated. In all cases, the form factors exhibit a linear dependence on the time interval.}
\label{Fig:dependence of formfactor on Q2}
\end{figure*}

In the data analysis we take a look at the $t$-dependence and the $t_f-t$ dependence of $\Gamma_{\phi\gamma\to G}^{(3)}$. For a fixed $\Delta t=t_f-t$ ($\Delta t/a_t=1$ for example), the form factor $E_1(Q^2,t)$ derived from the ratio function $R^{\mu i}(t,\Delta t)$ in Eq.~(\ref{Eq:ratio}) for a given $Q^2$ tends to a plateau for large $t$. However, for a fixed $t$ where the $\phi$ contribution to the three-point function dominates, $E_1(Q^2,\Delta t)$ derived from $R^{\mu i}(t,\Delta t)$ exhibits a clear linear behavior in $\Delta t$, as shown in the Fig.~\ref{Fig:dependence of formfactor on Q2}. This is unexpected because we do not seen this phenomenon in the $J/\psi\to \gamma G$ case. The only difference in the two cases is the change from charm quark to strange quark. It is known that quarks can propagate forward and backward in time, and the backward propagation is a relativistic effect and is suppressed by the quark mass for a massive quark. So the relativistic effect of strange quarks is much more pronounced over that of charm quarks, since charm quark is much heavier than strange quark. This relativistic effect of strange quarks may induce the propagation of a $s\bar{s}$ scalar meson that mixes with the scalar glueball.
\begin{figure}[h]
\centering
    \includegraphics[scale=0.2]{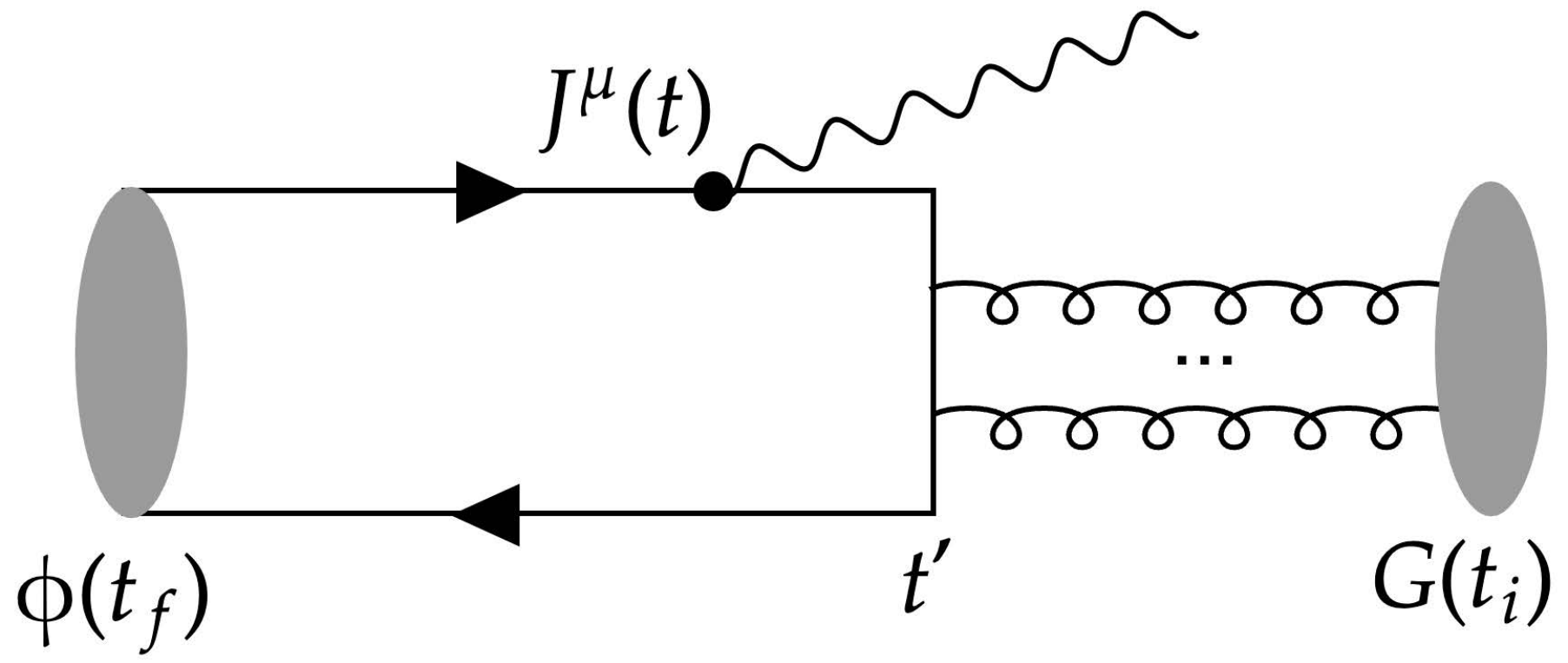}
\caption{Schematic diagram of $G$ and $s\bar{s}$ mixing.}
\label{Fig:App-skeleton}
\end{figure}

In order to see this glueball ($G$)-$s\bar{s}$ mixing effects explicitly, we make discussions in a two-state model involving the pure $s\bar{s}$ state $|s\bar{s}\rangle$ and the pure scalar glueball state $|G\rangle$ following the strategy in Refs.~\cite{Lee:1999kv,McNeile:2000xx,McNeile:2002az,McNeile:2002fh,Zhang:2021xvl}. The two states are orthogonal and satisfy the normalization conditions $\langle s\bar{s}|s\bar{s}\rangle =\langle G|G\rangle=1$. Let $|s\bar{s}\rangle$ be $(1,0)^T$ and $|G\rangle$ be $(0,1)^T$, then the Hamiltonian of the system in its rest frame can be expressed by
\begin{equation}
    \hat{H}=\left(
    \begin{array}{cc}
        E_s & x\\
         x  & E_G
    \end{array}
    \right),
\end{equation}
where $E_s$ and $E_G$ are the energies of $|s\bar{s}\rangle$ and $|g\rangle$, respectively, and $x$ is actually the transition amplitude $x=\langle s\bar{s}|\hat{H}|g\rangle$. If $x$ and the energy difference $\Delta=E_s-E_G$ are much smaller than $E_s$ and $E_G$, then it is easy to verify that, on a lattice of a temporal lattice spacing $a_t$, the transfer matrix $\hat{T}(a_t)=\exp(-a_t\hat{H})$ between two timeslices reads
\begin{eqnarray}\label{eq:App-transfer}
        \hat{T}(a_t)&=&e^{-a_t E} \left(
    \begin{array}{cc}
        e^{-a_t\Delta/2} & -a_t x\\
         -a_t x  & e^{a_t \Delta/2}
    \end{array}
    \right)\nonumber\\
    &=& \left(
    \begin{array}{cc}
        e^{-E_s a_t} & 0\\
         0  & e^{-E_G a_t}
    \end{array}
    \right)+e^{-a_t E}\left(
    \begin{array}{cc}
           0    & -a_t x\\
         -a_t x  & 0
    \end{array}
    \right)\nonumber\\
    &\equiv& \hat{T}_0(a_t)+\hat{T}'(a_t)
\end{eqnarray}
where $E=(E_s+E_G)/2$. 

Now let us start with the three-point function in Eq.~(\ref{Eq:Corr3_phi}). The relativistic effect of the strange quark makes it propagate forward and backward in time. As shown in Fig.~\ref{Fig:App-skeleton}, this effect can develop a scalar $s\bar{s}$ meson that propagate in the time range from the timeslice $t_i$ where the glueball operator $\mathcal{O}_G(t_i)$ is placed, to the timeslice $t$ where the EM current resides. Due to the same quantum number, the transition between the $s\bar{s}$ state and scalar glueball state takes place at any timeslice $t'\in [t,t_i]$ and is indicated in Fig.~\ref{Fig:App-skeleton} by a $G-s\bar{s}$ vertex. We ignore temporarily the momentum labels and spatial indices for simplicity. The three-point function $\Gamma^{(3)}(t_i=0,t_f;t)$ for $G\to\gamma\phi$ can be expressed as
\begin{widetext}
\begin{eqnarray}
    \Gamma^{(3)}(0,t_f;t)&=&\langle \Omega|\mathcal{O}_\phi(t_f)J(t)\mathcal{O}_G^\dagger(0)|\Omega\rangle = \langle\Omega|\mathcal{O}(0)e^{-\hat{H}_\phi(t_f-t)}J_{\text{em}}(0)[\hat{T}(a_t)]^{t/a_t}\mathcal{O}_G^\dagger(0)|\Omega\rangle\nonumber\\
    &=&\sum\limits_{\alpha\beta} \langle\Omega|\mathcal{O}_\phi(0)e^{-\hat{H}_\phi(t_f-t)}J_{\text{em}}(0)|\alpha\rangle\langle\alpha|[\hat{T}(a_t)]^{t/a_t}|\beta\rangle\langle\beta|\mathcal{O}_G^\dagger(0)|\Omega\rangle\nonumber\\
    &\approx&\sum\limits_{\alpha\beta} \langle\Omega|\mathcal{O}_\phi(0)e^{-\hat{H}_\phi(t_f-t)}J_{\text{em}}(0)|\alpha\rangle\langle\alpha|[\hat{T}_0]^{t/a_t}|\beta\rangle\langle\beta|\mathcal{O}_G^\dagger(0)|\Omega\rangle\nonumber\\
    &+&\sum\limits_{\alpha\beta}\sum\limits_{t'=0}^{t-a_t}\langle\Omega|\mathcal{O}_\phi(0)e^{-\hat{H}_\phi(t_f-t)}J_{\text{em}}(0)|\alpha\rangle\langle\alpha|[\hat{T}_0]^{(t-t')/a_t-1}\hat{T}'[\hat{T}_0]^{t'/a_t} |\beta\rangle\langle\beta|\mathcal{O}_G^\dagger(0)|\Omega\rangle,
\end{eqnarray}
where we only keep the leading terms of $\hat{T}'$ and the summations are over $|g\rangle$ and $|s\bar{s}\rangle$ states. If the approximation $\langle \Omega|\mathcal{O}_G|s\bar{s}\rangle\approx 0$ is assumed, then using Eq.~(\ref{eq:App-transfer}) and inserting the intermediate states between $\mathcal{O}_\phi$ and $e^{-\hat{H}_\phi(t_f-t)}$, one has 
\begin{eqnarray}
    \Gamma^{(3)}(0,t_f;t) &\approx& Z_\phi Z_G^* e^{-E_\phi(t_f-t)}\langle \phi|J_{\text{em}}(0)|g\rangle e^{-E_G t}\nonumber\\
    &-&Z_\phi Z_G^* e^{-E_\phi(t_f-t)}\sum\limits_{t'=0}^{t-a_t}\langle\phi|J_{\text{em}}(0)|s\bar{s}\rangle e^{-E_s (t-t'-1)} x a_t e^{-E a_t} e^{-E_G t'}.
\end{eqnarray}
for $(t_f-t)/a_t\gg 1$ (note that the operator $\mathcal{O}_G$ is optimized to couple predominantly to the ground glueball state $|g\rangle$), where $Z_\phi=\langle \Omega|\mathcal{O}_\phi(0)|\phi\rangle$ and $Z_G=\langle \Omega|\mathcal{O}_G(0)|G\rangle$ are defined.
It is easy to verify that 
\begin{equation}
    \sum\limits_{t'=0}^{t-a_t} e^{-E_s (t-t'-1)} xa_t e^{-E a_t} e^{-E_G t'} = a_t x e^{-Et} \frac{\sinh(t\Delta/2)}{\sinh(\Delta/2)}\approx xt\left(1+\frac{1}{24}(t\Delta)^2+\ldots\right) e^{-Et}.
\end{equation}
\end{widetext}
Therefore, for $E_s\approx E_G\approx E$ one has 
\begin{eqnarray}
    \Gamma^{(3)}(0,t_f;t) &\approx& Z_\phi Z_G^* e^{-E_\phi(t_f-t)}e^{-E t}\nonumber\\
    &\times& ( \langle \phi|J_{\text{em}}(0)|G\rangle +xt \langle \phi|J_{\text{em}}(0)|s\bar{s}\rangle), 
\end{eqnarray}
such that $E_1(Q^2)$ derived from Eq.~(\ref{Eq:ratio}) shows a linear dependence on $\Delta t=t_f-t$ for a fixed $t$. Practically, we first obtain $R^{\mu i}_{G\to \gamma \phi}(Q^{2},t,\Delta t)$ at different $\Delta t$ using Eq.~(\ref{Eq:ratio}), and then perform a fit to $E_{1}(Q^{2},\Delta t)$ by a linear function form to determine $E_1(Q^2)$. Fig.~\ref{Fig:dependence of formfactor on Q2} shows the $\Delta t$ behavior of $E_1(Q^2,\Delta t)$ at $Q^2=0.4390~\text{GeV}^2$ at $\beta=2.4$. It is neatly described by a linear function in $\Delta t$
\begin{figure}[t]
\centering
    \includegraphics[scale=0.48]{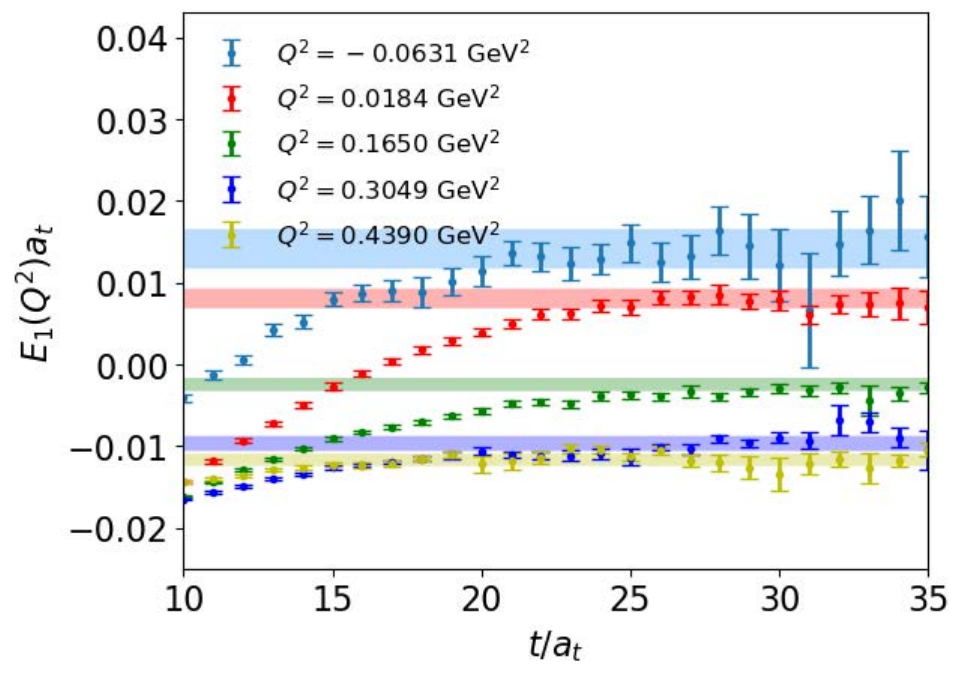}
\caption{Fitting of the form factor for $G \to \gamma \phi$. The figure shows the fitting results of $E_1(Q^2, t)$ at different $Q^2$ for $\Delta t = 0$ on the lattice with $\beta = 2.4$.}
\label{Fig:Change of form factor}
\end{figure}
\begin{equation}
E_1(Q^2,\Delta t)= E_1(Q^2)+c \Delta t
\end{equation}
with $E_1(Q^2)=-0.0502(25)~\text{GeV}$ and $c=-0.1928(55)~\text{GeV}^2$. We find that the value of $E_1(Q^2)$ matches the value at $E_1(Q^{2},\Delta t=0)$ within the error. In this manner, we obtain the results for the form factor $E_{1}(Q^{2})$, as shown in Fig.~\ref{Fig:Change of form factor}. 

\begin{figure}[t!]
    \centering
    \includegraphics[scale=0.48]{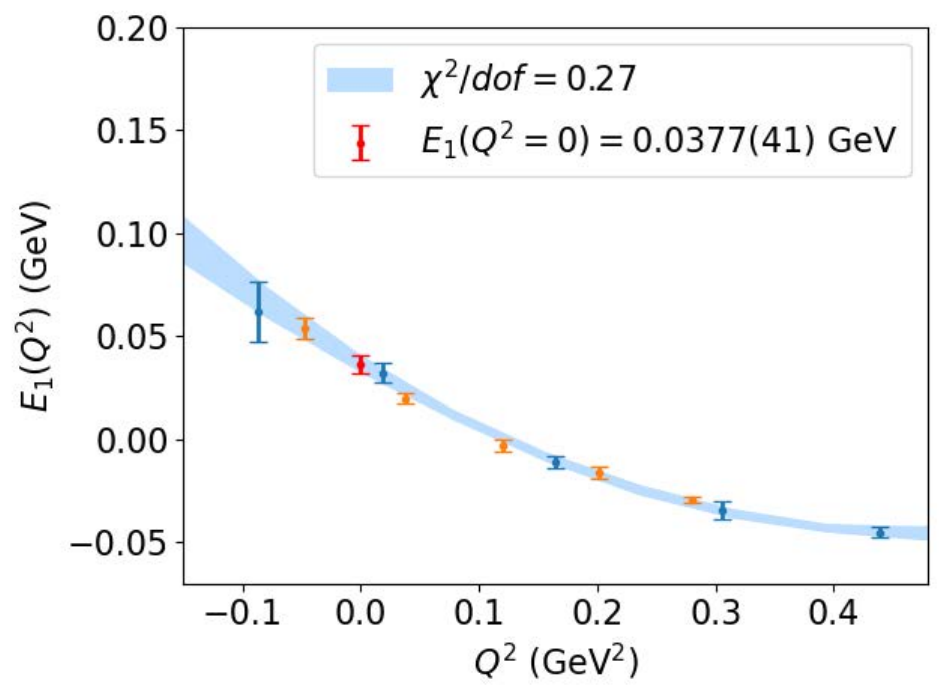}
    \includegraphics[scale=0.48]{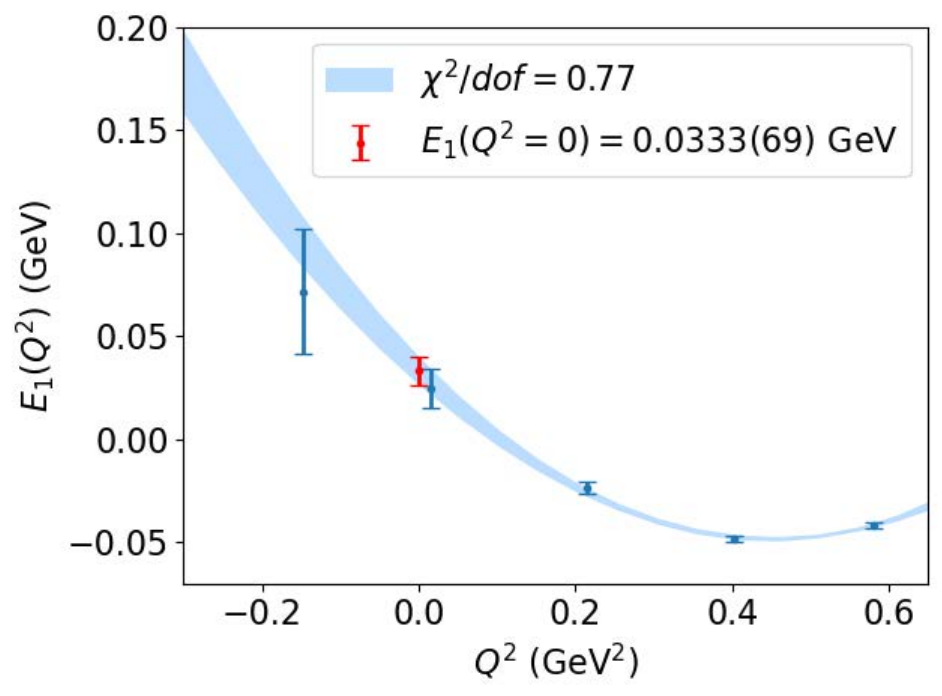}
    \includegraphics[scale=0.48]{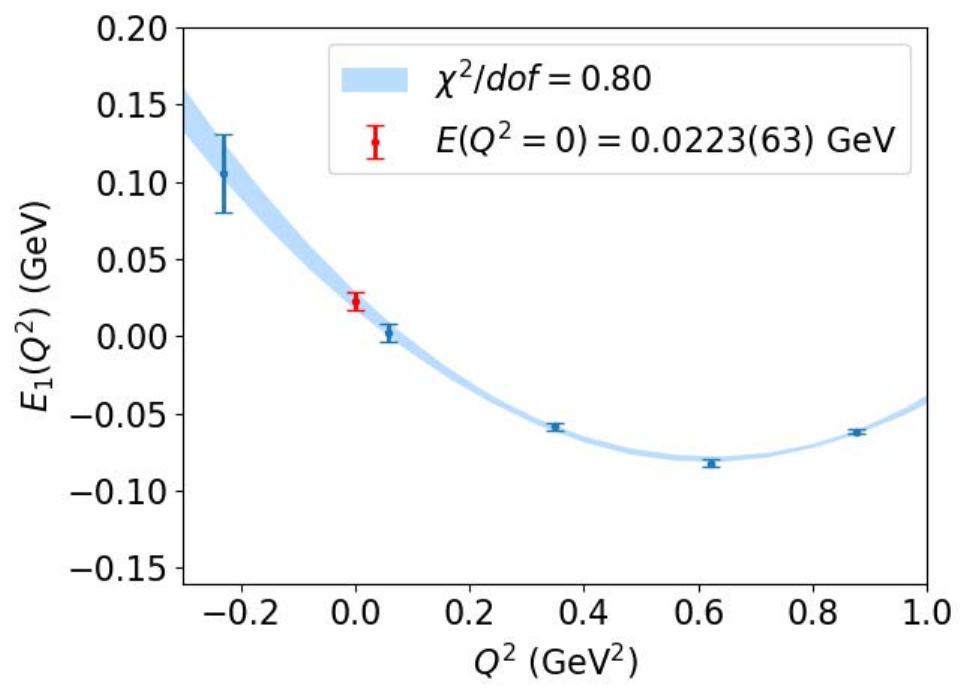}
\caption{Extrapolation of $E_1(Q^2)$ to the $Q^2 = 0$ for the $G \to \gamma \phi$ process. The results from top to bottom correspond to $\beta=2.4$, $\beta=2.8$, and $\beta=3.0$. For the ensemble with $\beta=2.4$, measurements were performed on two different lattice sizes, $12^3\times 192$ and $16^3 \times 192$. Our results indicate that finite volume effects are not significant.}
\label{Fig:form factors of scalar glueball to phi}
\end{figure}
\begin{figure}[ht!]
    \centering
    \includegraphics[scale=0.48]{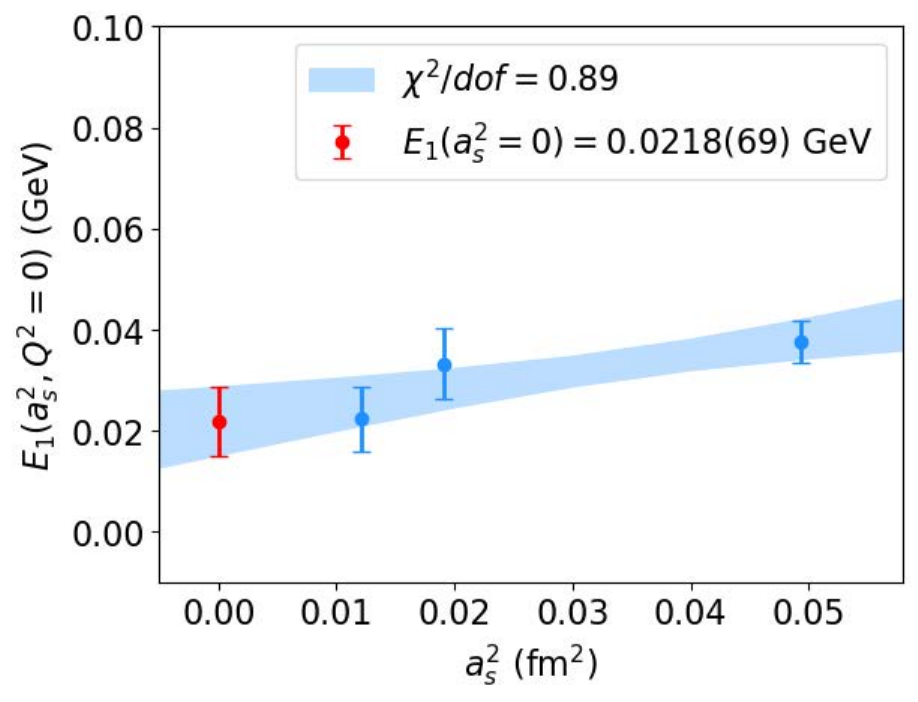}
\caption{The continuum limit extrapolation of the form factors $E_1$ for the $G \to \gamma \phi$ process is carried out at three different lattice spacings using linear extrapolation.}
\label{Fig:Linear extrapolation of form factor}
\end{figure}

The on-shell form factor $E_1(Q^2=0)$ is required to predict the partial decay width of $G\to\gamma \phi$. The interpolation to $Q^2=0$ is performed using the polynomial function form in Eq.~(\ref{eq:interpolation}). As shown in Fig.~\ref{Fig:form factors of scalar glueball to phi} by shaded curves, this function form describes the data very well for $\beta=2.4,2.8,3.0$. With the values of $E_1(Q^2)$ at the three different lattice spacings $a_s$, we carry out a linear extrapolation in $a_s^2$ to the continuum limit and obtain the final value for the form factor $E_1(0)$ and the fitting situation is shown in Fig.~\ref{Fig:Linear extrapolation of form factor},
\begin{equation}
    E_{1}(0)^{\text{cont.}} = 0.0218(69)~\text{GeV},
\end{equation}
which gives the prediction of the partial decay width of the process $G\to\gamma\phi$
\begin{equation}
    \Gamma_{G\rightarrow\gamma\phi} = 0.074(47)~\text{keV},
\end{equation}
through the second relation in Eq.~(\ref{Eq:decay width}).

\section{DISCUSSION}
\label{DISCUSSION}
Glueballs, as bound states of gluons, are well defined objects in pure Yang-Mills theories. Their existence has been confirmed by previous lattice QCD calculations in the quenched approximation~\cite{morningstar1999glueball,chen2006glueball,Athenodorou:2020ani}. 
There are also some lattice QCD calculations with dynamical quarks using the gluonic operators to predict the glueball masses and having obtained results consistent with those from the quenched approximation~\cite{Richards:2010ck,gregory2012towards,bali2000static,sun2018glueball,chen2023glueballs}. Strictly speaking, the definition of glueballs in full QCD is not so conceptually rigorous as that in the quenched approximation owing to the gluon-quark transition. With this in mind, the phenomenological studies of hadrons usually conjecture that glueballs do exist (at least as a degree of freedom) and a physical meson state can be a $q\bar{q}$ meson, a multiquark state, a glueball, or an admixture of them. In this sense, the glueball properties derived in the quenched appproximation serve as theoretical inputs to the phenomenological interpretation of the nature of a meson, especially to identify its glueball components. This is the motivation of our discussion in the following.


We have obtained the EM form factors $E_1(0)$ for the processes $J/\psi\to\gamma G$ and $G\to \gamma\phi$ in the continuum limit. $E_1(0)$ is actually the effective coupling $g_\text{eff}$ of the effective Lagrangian for the decay processes above
\begin{equation}
    \mathcal{L}_\text{eff}=e Q_q g_\text{eff} G V_\mu A^\mu,
\end{equation}
where $V_\mu, G$ are the fields for the vector meson ($J/\psi$ or $\phi$ in this work) and the scalar glueball, respectively, $A_\mu$ is the electromagnetic field, $-e$ is the electric charge of electron, $Q_q$ is the electric charge of the quark in the vector meson, and $g_\text{eff}$ describes the $V\leftrightarrow G$ transition. Considering the expressions in Eq.~(\ref{Eq:decay width}), one can see $g_\text{eff}=E_1(0)$. If we introduce a dimensionless coupling constant $g_{G\phi}=E_1(0)/m_G$ for $G\to\gamma\phi$ and a coupling constant $g_{G J/\psi}=E_1(0)/m_{J/\psi}$ for $J/\psi \to \gamma G$, then we get
\begin{equation}
    g_{G\phi}/g_{GJ/\psi}=0.73(41)\sim \mathcal{O}(1)
\end{equation}
for $m_G\approx 1.635(62)~\text{GeV}$. This signals the insensitivity of $g_{GV}$ to quark masses. 

The partial decay width of $J/\psi\to \gamma G$ is predicted to be $0.578(86)~\text{keV}$, which gives the branching fraction 
\begin{equation}
    \text{Br}(J/\psi\to\gamma G)=6.2(9)\times 10^{-3},
\end{equation}
when the $J/\psi$ total width $\Gamma=92.6(1.7)~\text{keV}$~\cite{particle2022review} is used. These results are slightly larger than but qualitatively compatible with the previous lattice results $0.35(8)~\text{keV}$ and $3.8(9)\times 10^{-3}$~\cite{gui2013scalar}. As for the scalar glueball candidates $f_0(1710)$ and $f_0(1500)$, summing up the PDG data of the branching fractions of $J/\psi\to \gamma f_0(1710)\to \gamma (K\bar{K},\pi\pi, \eta\eta,\omega\omega,\omega\phi)$~\cite{particle2022review} gives the lower bound $\mathrm{Br}(J/\psi \to \gamma f_0(1710)) > 2.1 \times 10^{-3}$, while the lower bound for $f_0(1500)$ is $\mathrm{Br}(J/\psi\to \gamma f_0(1500)) > 1.9\times 10^{-4}$ by summing up the branching fractions of $J/\psi\to \gamma f_0(1500)\to \gamma (\pi\pi,\eta\eta,K\bar{K})$. Obviously, the production rate of $f_0(1710)$ in the $J/\psi$ radiative decay is one order of magnitude larger than that of $f_0(1500)$ are consistent with the production rate of the scalar glueball. On the other hand, BESII and BESIII have performed the partial wave analysis (PWA) of the processes $J/\psi\to \gamma X\to \gamma \pi\pi$~\cite{Ablikim:2006db}, $\gamma \eta\eta$~\cite{BESIII:2013qqz}, $\gamma K_S K_S$~\cite{BESIII:2018ubj}, and found that in each process, $f_0(1710)$ is produced much more than $f_0(1500)$. These observations support $f_0(1710)$ to be predominantly a scalar glueball state or have a large component of the scalar glueball. Through a coupled channel analysis to $J/\psi\to \gamma(\pi\pi,K\bar{K},\eta\eta,\phi\omega)$ processes and considering the octet-singlet mixings of scalar mesons, Klempt {\it et al.} claim that there should be a glueball state with the parameters $(m,\Gamma)=(1865\pm 25_{-30}^{+10},370\pm50_{-20}^{+30})~\text{MeV}$, and its observed yield in radiative $J/\psi$ decays is $5.8(1.0)\times 10^{-3}$~\cite{Klempt:2021wpg,Sarantsev:2021ein,Gross:2022hyw}. 

Now we discuss the physical significance of the partial width $\Gamma(G\to\gamma\phi)=0.074(47)~\text{keV}$ predicted in this study. Since the width of a scalar glueball is expected to be $\mathcal{O}(100)~\text{MeV}$, this decay process has a branching fraction as small as $\mathcal{O}(10^{-6})$ and is hard to be detected directly. However, this branching fraction is helpful for experiments to judge the property of an intermediate scalar meson under some circumstances. Especially for the decay processes $J/\psi\to\gamma X\to \gamma \gamma\phi$ through scalar resonances, if $X$ is a glueball state, then the combined branching fraction is estimated to be 
\begin{equation}
    \mathrm{Br}(J/\psi\to\gamma G, G\to\gamma\phi)\sim \mathcal{O}(10^{-9}),
\end{equation} 
so the process $J/\psi\to\gamma\gamma\phi$ through the scalar glueball $G$ is hardly observed by BESIII even with the very large sample of $10^{10}$ $J/\psi$ events~\cite{BESIII:2021cxx}. Recently, BESIII reported the partial wave analysis results of $J/\psi\to\gamma\gamma\phi$~\cite{BESIII:2024ein}. There is no evidence for $f_0(1500)$ and $f_0(1710)$ in the $\gamma\phi$ system. The only scalar ($0^{++})$ component of $X$ is $f_0(2200)$ of a statistical significance greater than $5\sigma$ and a branching fraction $\mathrm{Br}(J/\psi\to\gamma f_0(2200), f_0(2200)\to \gamma\phi)=2.0(4)\times 10^{-7}$. So $f_0(2200)$ can be excluded from the scalar glueball candidates. There exist a few phenomenological studies on the radiative decays of the scalar glueball~\cite{Cotanch:2004py,hechenbergerRadiativeMesonGlueball2023}: Ref.~\cite{Cotanch:2004py} uses the vector meson dominance and the Regge/pomeron phenomenology and gives the partial width a fairly large value of $454~\text{keV}$. A recent study~\cite{hechenbergerRadiativeMesonGlueball2023} using the Witten-Sakai-Sugimoto model predicts this partial width is roughly a few tens of $\text{keV}$. In Ref.~\cite{Cotanch:2004py}, the effective coupling $g_{G\phi\gamma}$ for $G\to\gamma\phi$ is related to the effective coupling $g_{G\gamma\gamma}$ for the two-photon decay $G\to\gamma\gamma$ by $g_{G\phi\gamma}/g_{G\gamma\gamma}\approx 5.36$ based on the vector meson dominance, using the decay constants of the vector mesons $\rho$, $\omega$ and $\phi$ derived from PDG~\cite{particle2022review}. If this is actually the case, then one has~\cite{Cotanch:2004py} 
\begin{equation}
    \frac{\Gamma(G\to\gamma\phi)}{\Gamma(G\to \gamma\gamma)}=\frac{1}{2\pi\alpha}\left(\frac{g_{G\phi\gamma}}{g_{G\gamma\gamma}}\right)^2 \left(1-\frac{m_\phi^2}{m_G^2}\right)^3\approx 143
\end{equation}
for $m_G\approx 1635~\text{MeV}$. Then our results of $\Gamma(G\to\gamma\phi)$ implies that 
\begin{equation}
    \Gamma(G\to\gamma\gamma)\approx 0.52(33)~\text{eV},
\end{equation}
which provides a quantitative estimate of the stickiness of the scalar glueball~\cite{Chanowitz:1984cb,Crede:2008vw}
\begin{equation}
    S(G)=C \left(\frac{m_G}{q_\gamma}\right) \frac{\Gamma(J/\psi\to\gamma G)}{\Gamma(G\to \gamma\gamma)}\sim \mathcal{O}(10^4),
\end{equation}
where $C\approx 17.7$ is taken to make the stickiness of $f_2(1270)$ to be $S(f_2(1270))=1$ using the PDG data~\cite{particle2022review}, $q_\gamma=(m_{J/\psi}^2-m_G^2)/(2m_{J/\psi})$ is the momentum of the photon in the process $J/\psi\to\gamma G$ (in the rest frame of $J/\psi$). 
This value can be use as a reference to identify a pure glueball state or estimate the glueball component of a meson in the future experiments. The phenomenological study based on the non-relativistic gluon bound state model also predicts the partial width $\Gamma(G\to \gamma \gamma)$ to have similar magnitude~\cite{Kada:1988rs}.

\section{SUMMARY}
\label{SUMMARY}
We perform the first lattice QCD study on the radiative decay of the scalar glueball $G\to\gamma\phi$ in the quenched approximation. The calculations are conducted on three large gauge ensembles on anisotropic lattices with the spatial lattice spacings $a_s$ ranging from $0.222(2)~\text{fm}$ to $0.110(1)~\text{fm}$, which enables us to do a reliable continuum extrapolation. 

We first revisit the radiative $J/\psi$ decay to the scalar glueball and derive the electromagnetic form factor $E_1(0)=0.0677(48)~\text{GeV}$ in the continuum limit, which gives the partial decay width $\Gamma(J/\psi\to\gamma G)=0.578(86)~\text{keV}$ and the branching fraction $\mathrm{Br}(J/\psi\to\gamma G)=6.2(9)\times 10^{-3}$ when $m_G\approx 1.635(62)~\text{GeV}$ and the total width $\Gamma=92.6(1.7)~\text{keV}$ of $J/\psi$ are used. Comparing to the previous lattice results $0.35(8)~\text{keV}$ and $3.8(9)\times 10^{-3}$~\cite{gui2013scalar}, this work includes the data at finer lattice spacing and uses wall source propagators, which effectively improve the signal. This provides more solid support for $f_0(1710)$ being a candidate for the scalar glueball or having a large component of it. 

By calculating the on-shell form factor $E_1(0)=0.0218(69)~\text{GeV}$, we predict the partial decay width of $G\to \gamma\phi$ to be $\Gamma(G\to \gamma\phi)=0.074(47)~\text{keV}$. Considering the glueball width of $\mathcal{O}(100)~\text{MeV}$, this tiny value implies that the process $(J/\psi\to\gamma G, G\to \gamma\phi)$ is hardly observed even for the BESIII Collaboration who possesses a large sample of $\mathcal{O}(10^{10})$ events. Recently, BESIII reported the first partial wave analysis of the process $(J/\psi\to\gamma X$, $X\to \gamma\phi)$ where only one scalar component $f_0(2200)$ of $X$ is observed. By using the ratio of the $G-\phi-\gamma$ coupling to the $G-\gamma-\gamma$ coupling derived from the vector meson dominance model and our value of $\Gamma(G\to \gamma\phi)$, we estimate the two-photon decay width of the scalar glueball to be $\Gamma(G\to \gamma\gamma)\approx 0.53(46)~\text{eV}$. This provides a new quantitative value $S(G)\sim
\mathcal{O}(10^4)$ for the stickiness of the pure scalar glueball. 

\begin{acknowledgments}
This work is supported by the National Natural Science Foundation of China (NSFC) under Grants No. 12175063, No.\ 12175073, No.\ 12222503, No.\ 11935017, No.\ 12293060, No.\ 12293062, No.\ 12293065, No.\ 12070131001 (CRC 110 by DFG and NSFC)). JL is also supported by the Natural Science Foundation of Basic and Applied Basic Research of Guangdong Province under Grants No.\ 2023A1515012712. CY also acknowledges the support by the National Key Research and Development Program of China (No.\ 2020YFA0406400) and the Strategic Priority Research Program of Chinese Academy of Sciences (No.\ XDB34030302). LG also acknowledges the support by the Hunan Provincial Natural Science Foundation (No.\ 2023JJ30380). WQ also acknowledges the support by the Hunan Provincial Natural Science Foundation (No.\ 2024JJ6300) and the Scientific Research Fund of Hunan Provincial Education Department (No.\ 22B0044). The numerical calculations are carried out on the GPU cluster at Hunan Normal University. Our matrix inversion code is based on QUDA libraries~\cite{clark2010solving} and the fitting code is based on lsqfit~\cite{lepage2002constrained}.
\end{acknowledgments}

\bibliography{myref}
\end{document}